\documentclass[iop]{emulateapj-rtx4} 
\shortauthors{Sekanina}
\shorttitle{Outburst and Splitting of Comet C/2015 ER$_{61}$}
\slugcomment{Version \today }

\begin{document}
\title{MAJOR OUTBURST AND SPLITTING OF LONG-PERIOD COMET C/2015 ER$_{61}$ (PAN-STARRS)}
\author{Zdenek Sekanina}
\affil{Jet Propulsion Laboratory, California Institute of Technology,
  4800 Oak Grove Drive, Pasadena, CA 91109, U.S.A.}
\email{Zdenek.Sekanina@jpl.nasa.gov}

\begin{abstract}
In early April 2017, five weeks before passing through perihelion, the dust-poor
comet C/2015 ER$_{61}$ was observed to undergo a major outburst, during which its
intrinsic brightness increased briefly by about 2~mag.  Evidence is presented
in this paper to suggest that the flare-up was in all likelihood a product of an
event of nuclear fragmentation that gave birth to a companion nucleus, which
was first detected more than nine weeks later, on June~11, and remained under
observation for 19~days.  The companion is found to have separated from the
parent nucleus with a velocity of less~than~1~m~s$^{-1}$; subsequently it
was subjected to a nongravitational deceleration of 14~units of 10$^{-5}$\,the
Sun's gravitational acceleration.  The companion's overall dimensions can from
this value be estimated at less than one hundred meters across.  Unlike the
primary nucleus, the companion displayed considerable fluctuations in brightness
on time scales spanning at least three orders of magnitude, from a fraction of
one hour to weeks.  The long temporal gap between the companion's birth and
first detection appears to be a corollary of the brightness variability.  Evidence
from the light curve suggests that the companion became detectable only because
of large amounts of debris that it started shedding as well as increased activity
from areas newly exposed by progressive fragmentation, the same process that
also resulted in the companion's impending disintegration by the end of June or
in early July.
\end{abstract}


\section{Introduction}
First spotted in several images taken under auspices of the Pan-STARRS Project
with the 180-cm f/2.7 Ritchey-Chr\'etien reflector on Haleakal$\stackrel{\_}{\rm
a}$, Maui, on March 14--15, 2015 (Gibson et al.\ 2015),\footnote{See also {\it
Minor Planet Circulars Supplement\/}, MPS 592555.} this object was of star-like
appearance and classified as an asteroid 2015~ER$_{61}$ by the {\it Minor Planet
Center\/} (2015), even though it was almost 9~AU from the Sun and approaching it
along a nearly-parabolic orbit with a perihelion distance comparable~to 1~AU.
The object was subsequently detected in~\mbox{several} earlier Pan-STARRS images
from January and \mbox{February} 2015 (Primak et al.\ 2016; Gibson et al.\ 2016).
When signs of cometary activity were noticed by H.~Sato in late December 2015 and
confirmed by K.~S\'arneczky in January 2016, the object was re-classified as a
long-period comet (Green 2016). However,{\vspace{-0.04cm}} an original barycentric
reciprocal~semimajor~axis~of~\mbox{$+0.001389 \pm 0.0000005$ AU$ ^{-1}$} and
the corresponding orbital period of \mbox{$19\,320\pm\!10$ yr} (Nakano 2017)
demonstrate conclusively that the comet has not arrived from the Oort Cloud.

A typical dust-poor comet by its appearance (see Section~2.4), C/2015 ER$_{61}$ was
brightening steadily along its inbound orbit until a major outburst was reported
in progress on April~4, 2017 (King 2017), the brightness having eventually risen
to magnitude 6 (Green 2017a) before the flare-up began to subside a few days later.

Two months after this event, on June 13, a double nucleus was observed by
E.~Bryssinck (Green 2017b); the primary component, referred to hereafter as
nucleus A, was accompanied by a very faint companion B, of apparent magnitude
$\sim$16, located in the primary's coma some 0$^{\prime}\!$.2 away in an
essentially antisolar direction (\mbox{Hambsch} et al.\ 2017a; Masi \&~Schwartz
2017).  Subsequently, the companion was identified in two June~11 images, in
which it was still a bit fainter (Hambsch \& Bryssinck 2017).  The last of 52
astrometric observations of the companion come from June 30 (Hambsch et al.\
2017b),\footnote{All 52 available astrometric observations of the companion are
listed in {\it Minor Planet Circulars\/}~105299--105300.} by which time it faded
to magnitude 17.  The orbital arc thus covers 19~days, long enough to warrant an
in-depth investigation of the companion's motion.

Given the notoriety of temporal correlations between outbursts and nuclear
fragmentation of many comets (e.g., Sekanina 1982, 2010; Boehnhardt 2004), the
prime objective of this paper is the examination of a chance that the observed
outburst and nuclear duplicity of C/2015~ER$_{61}$ are, indeed, products of the
{\it same event\/}.  An important part of this task is the outburst's comprehensive
analysis, focused in particular on the best possible determination of the time
of its inception, a critical parameter for testing the significance of the
correlation between the outburst and the nucleus' breakup.

\section{Light Curve and the Onset of the Outburst}
To derive a dependable light curve of a comet is a difficult task, because the
procedure requires that two contradictory conditions be satisfied simultaneously:\
(i)~the temporal density of the data points be as high as possible and (ii)~the
data points be photometrically as homogeneous as possible.  This second
stipulation is equivalent to providing a {\it common magnitude scale\/} for all
data sources (data source = observer\,+\,telescope used).  An optimum solution
inevitably demands a delicate balance between the two constraints.

For sets of magnitudes of C/2015 ER$_{61}$, the second~condition could in practice
be accommodated, to a limited degree, by applying constant corrections to
the individual sources, which  may conveniently be determined graphically by
superposing transparencies with the light curves on top of one another until
an optimum match has been achieved among them (Section 2.2).

\subsection{Light Curve from CCD Observations}
A fairly coherent light curve was constructed by combining the sets of total
CCD magnitudes\footnote{These magnitudes, published by the {\it Minor Planet
Center\/}, are marked T in order to distinguish them from the so-called
nuclear magnitudes; see {\tt http://www.minorplanetcenter.net/db\_search}.}
reported~by~five Japanese observers.  This combined dataset offers one major
advantage:\ because of a very limited geographic distribution of the observers,
they imaged the comet at practically the same time of the day, to within an
hour or so of one another, so that the magnitudes they reported from the same
dates refer to virtually coincident times and provide an opportunity to derive
averaged systematic corrections between the observers' magnitude scales that
are unaffected by the comet's intrinsic brightness vari\-ations on the time
scales as short as $\sim$0.1~day.

Although three of the five observers began to record the magnitudes of
C/2015~ER$_{61}$ as early as February--April 2016, I examine their light curves
only from the second half of November 2016 on, still nearly six months before
perihelion, when the magnitude reports arrived in earnest after a gap of more
than four months, after the comet emerged from a conjunction with the Sun.  The
examined dataset was terminated in the second half of June 2017, some six weeks
after perihelion.  During the more than seven months covered, the five CCD
observers made a total of 54 determinations of the comet's brightness.  Of
primary interest was the period of three months, from early February 2017 to
perihelion, which covers the reported outburst.

The five observers are listed in Table 1, which includes their abbreviations
employed below, their locations and observatory codes, the telescopes used, the
number of observations made, and the number of temporally coincident observations
that are used below to derive the systematic magnitude corrections.  Marked in
the table is the reference observer, to whose magnitude scale the corrections for
the other observers are referred.

The self-explanatory Table 2 presents a list of the temporally coincident
observations by the five Japanese observers over a period of more than two
months approximately centered on the comet's perihelion time.  This information
can readily be employed to compute the magnitude offsets between any two among
the five observers, whose averaged values will become the adopted corrections.
To start with, let for a pair of observers, {\sf X}, {\sf Y}, for which
\mbox{{\sf X}\,=\,\{{\sf Ab},\,{\sf Ka},\,{\sf Oh},\,{\sf Se},\,{\sf Ta}\}},
\mbox{{\sf Y}\,=\,\{{\sf Ab},\,{\sf Ka},\,{\sf Oh},\,{\sf Se},\,{\sf Ta}\}},
and \mbox{{\sf X}\,$\neq$\,{\sf Y}}, the reported magnitudes at a given time
be, respectively, $H({\sf X})$ and $H({\sf Y})$, and let their difference,
\begin{equation}
\Re({\sf X},{\sf Y}) = H({\sf X}) - H({\sf Y}),
\end{equation}
be referred to as an offset of the magnitude reported by observer {\sf Y} from
the magnitude reported simultaneously by observer {\sf X}.  If there is a third
reporting observer, {\sf Z}, one can always write Equation~(1) as
\begin{eqnarray}
\Re({\sf X},{\sf Y}) & = & \Re({\sf X},{\sf Z}) - \Re({\sf Y},{\sf Z}) \nonumber \\
                     & = & \Re({\sf X},{\sf Z}) + \Re({\sf Z},{\sf Y}).
\end{eqnarray}
\begin{table}[t]
\vspace{-4.18cm}
\hspace{4.22cm}
\centerline{
\scalebox{1}{
\includegraphics{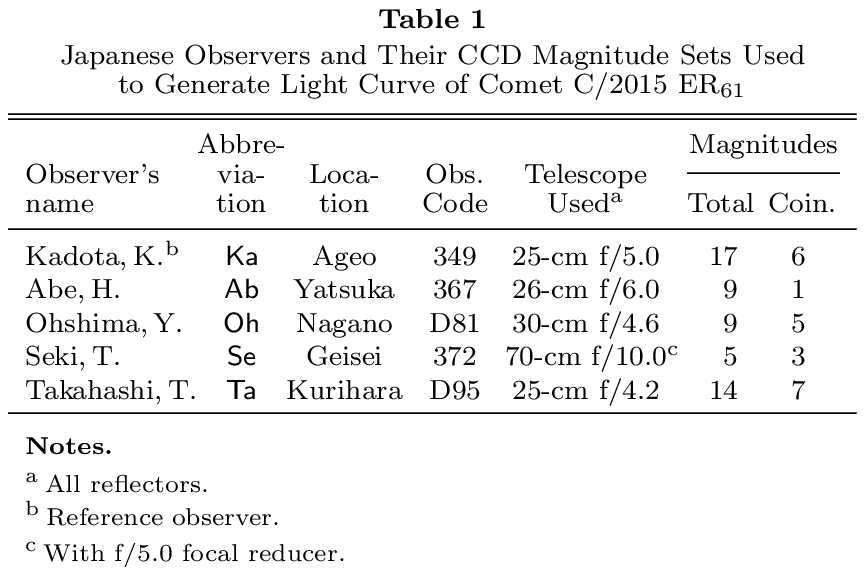}}}
\vspace{-19.55cm}
\end{table}

The purpose of introducing a reference observer, {\sf R}, is to formalize the
system of corrections among a number of observers:\ all magnitudes are
eventually referred to observer {\sf R} by an equation of type~(2).  It is
convenient to adopt a rule that {\sf R} be always the first argument of
$\Re$ [i.e., {\sf X} in Equation~(2)].  If so, the symbol for an offset of a
magnitude reported by observer {\sf Z} relative to a magnitude reported at
the same time by the reference observer {\sf R} can formally be simplified:
\begin{equation}
\Re({\sf R},{\sf Z}) \equiv \Re({\sf Z}),
\end{equation}
so that an offset of magnitudes by two non-reference observers, {\sf Y} and
{\sf Z}, can be expressed by
\begin{equation}
\Re({\sf Y},{\sf Z}) = \Re({\sf Z}) - \Re({\sf Y}).
\end{equation}
\begin{table}[b]
\vspace{-3.8cm}
\hspace{4.22cm}
\centerline{
\scalebox{1}{
\includegraphics{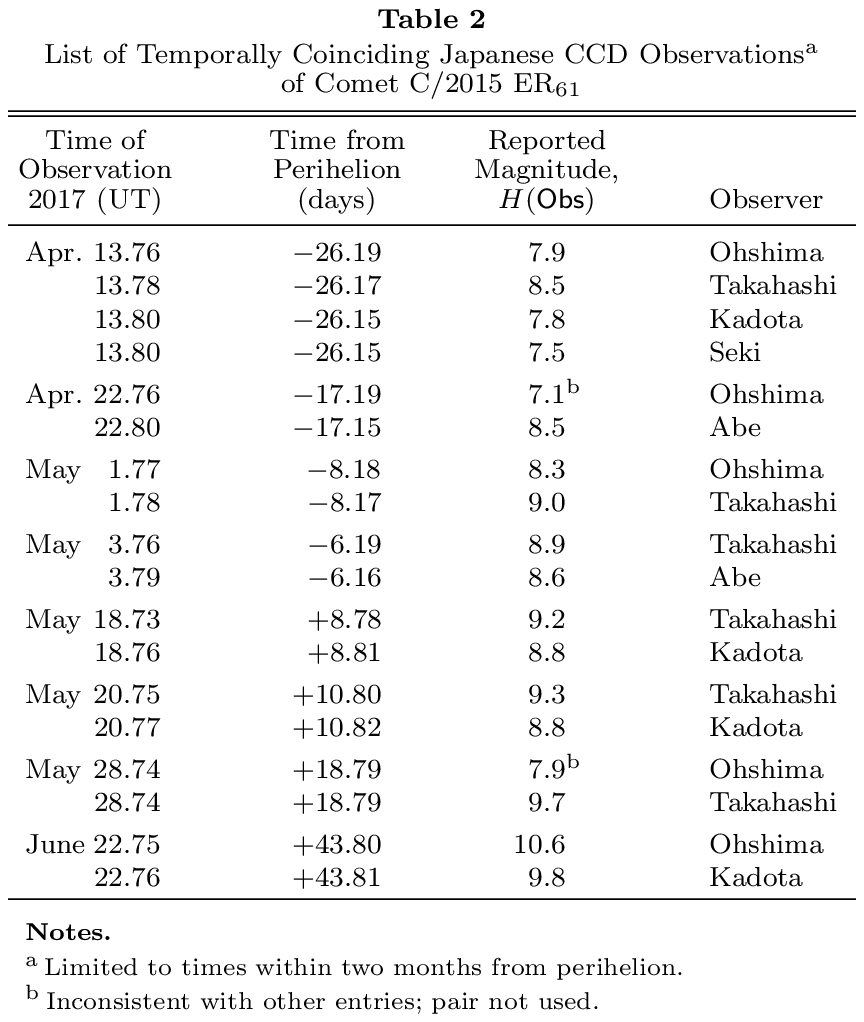}}}
\vspace{-15.5cm}
\end{table}

Before turning to Table 2, it is useful to recall that from magnitudes
reported by $n$ different observers for a given time one can derive $N$
independent magnitude offsets, where
\begin{equation}
N = {\textstyle \frac{1}{2}} n (n \!-\! 1),
\end{equation}
that is, one offset from magnitudes by two observers, three offsets from
magnitudes by three observers, etc.

The magnitude offsets generated from the data in Table~2 can be divided into
two groups:\ those including an observation by the reference observer {\sf Ka}
and those excluding it.  There is a total of 11 offsets.  Of these six include
{\sf Ka}:\ three from April~13 and one each from May~18, May 20, and June~22.
Among these, three are {\sf Ta} offsets, two {\sf Oh} offsets, and one {\sf Se}
offset.  Five offsets exclude {\sf Ka}:\ three from April~13 and one each from
May~1 and May~3.  Among these, two relate {\sf Oh} with {\sf Ta} and one relates
{\sf Ab} with {\sf Ta}, {\sf Oh} with {\sf Se}, and {\sf Se} with {\sf Ta}.

The $N_{\sf X}$ indidivual determinations of the offsets of the magnitude scale
of each observer, {\sf X}, relative to the reference observer are averaged in
order to derive an effective correction, $\langle \Re({\sf X}) \rangle$,
\begin{equation}
\langle \Re({\sf X}) \rangle = \frac{1}{N_{\sf X}} {\mbox{\raisebox{-0.3ex}{\Large
 \boldmath $\Sigma$}}}{\hspace{-0.35cm}}{\mbox{\raisebox{-1.5ex}{\tiny \sf (X)}}}
 \,\: \Re({\sf X}).
\end{equation}

The offsets that do not include an observation by the reference observer
have first to be converted to include {\sf Ka} using Equation~(4).  Of the
four non-reference observers in Table~1, it is noticed from Table~2 that
offsets of magnitudes by observer {\sf Ab} relative to the reference observer
are unavailable.  Offset $\Re({\sf Ab})$ should therefore be used as an
independent variable in a search for a minimum~sum of squares of residuals
for all available offsets $\Re({\sf X})$ of the non-reference observers,
\begin{equation}
{\mbox{\raisebox{-0.55ex}{\Large \boldmath $\Sigma$}}}{\hspace{-0.47cm}}{\mbox{
 \raisebox{-1.7ex}{\tiny \sf (X)}}} \, \left[\Re({\sf X}) \!-\!
 \langle \Re({\sf X}) \rangle \right]^2 = {\rm min}.
\end{equation}
\begin{table}[t]
\vspace{-4.2cm}
\hspace{4.22cm}
\centerline{
\scalebox{1}{
\includegraphics{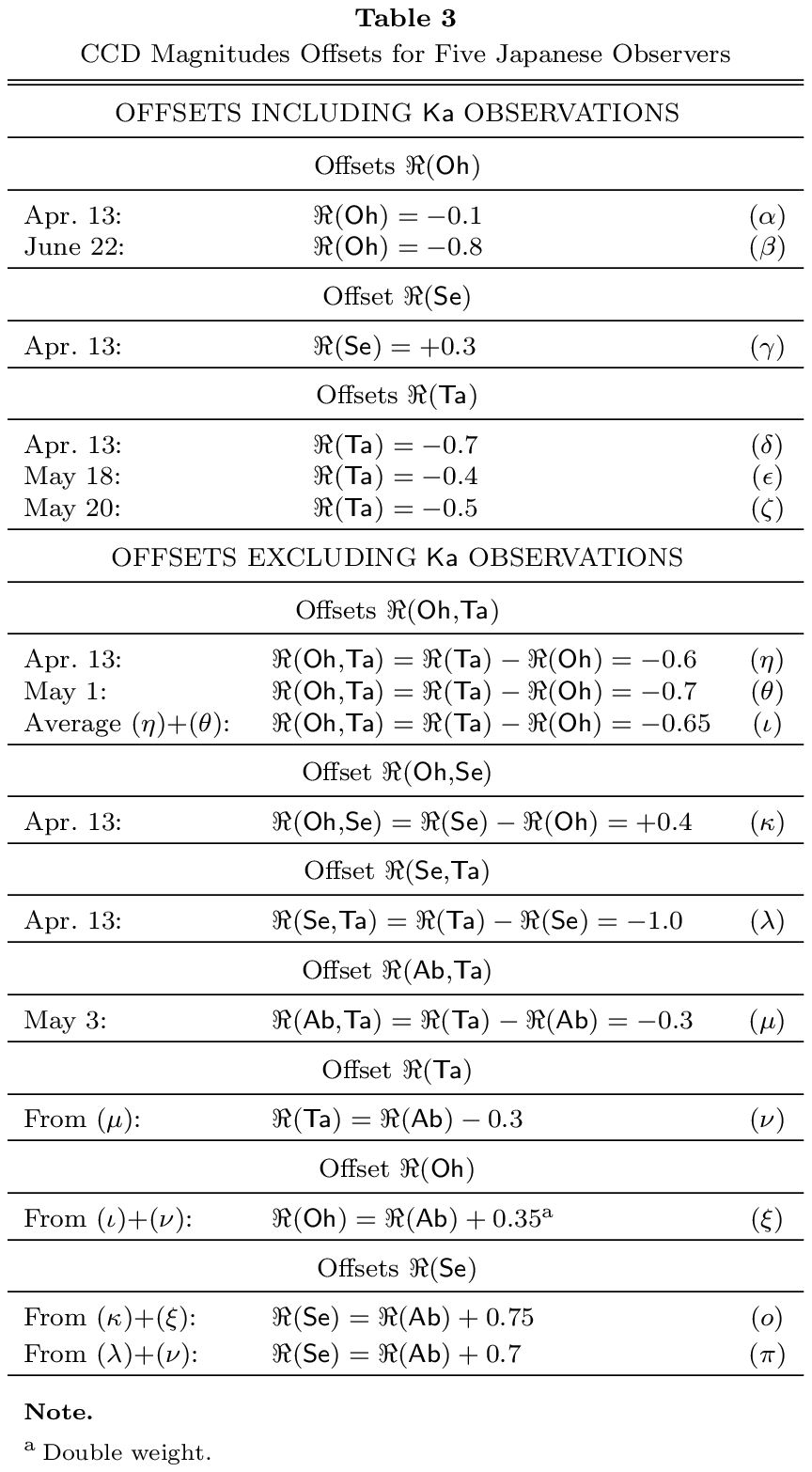}}}
\vspace{-9.5cm}
\end{table}

The procedure is apparent from Table 3.  Its upper part presents the six
offsets ($\alpha$) through ($\zeta$) that include {\sf Ka}, the lower part
converts the remaining offsets into the four conditions ($\nu$) through
($\pi$), with the condition ($\xi$) having a double weight because of the
averaging of the conditions ($\eta$) and ($\theta$) carried out for the
sake of convenience.  Altogether one has for each assumed $\Re({\sf Ab})$
four $\Re({\sf Oh})$ conditions, ($\alpha$), ($\beta$), and twice ($\xi$);
three $\Re({\sf Se})$ conditions, ($\gamma$), ($o$), and ($\pi$); and
four $\Re({\sf Ta})$ conditions, ($\delta$), ($\epsilon$), ($\zeta$), and
($\nu$).

A practical approach to the problem consists in choosing a value of $\Re({\sf
Ab})$; calculating $\Re({\sf Oh})$, $\Re({\sf Se})$, and $\Re({\sf Ta})$ from
($\nu$) through ($\pi$) in Table~3; averaging these offsets, as shown by
Equation~(6), to obtain the corrections $\langle \Re({\sf Oh}) \rangle$,
$\langle \Re({\sf Se}) \rangle$, and $\langle \Re({\sf Ta}) \rangle$,
respectively; calculating the 11 residuals; summing up their squares according
to Equation~(7); and repeating these steps until a minimum is found on the
curve of the squares of residuals.  The respective $\Re({\sf Ab})$ equals the
correction $\langle \Re({\sf Ab}) \rangle$ and determines the corrections
for the observers {\sf Oh}, {\sf Se}, and {\sf Ta} relative to {\sf Ka}.

Application of this approach leads to the following effective corrections
to convert the magnitude observations of the four observers to that of
Kadota:
\begin{eqnarray}
\langle \Re({\sf Ab}) \rangle = -0.52, \nonumber \\
\langle \Re({\sf Oh}) \rangle = -0.31, \nonumber \\
\langle \Re({\sf Se}) \rangle = +0.24, \nonumber \\
\langle \Re({\sf Ta}) \rangle = -0.61,\,
\end{eqnarray}
with the sum of squares or residuals equal to 0.4390~mag$^2$ and a mean residual
of $\pm$0.21~mag.  Referring the~data points by Abe, Ohshima, Seki, and Takahashi
to the magnitude scale of Kadota using the corrections in Equations~(8), the
resulting light curve is plotted in Figure~1. 

\begin{figure*}
\vspace{-5.3cm}
\hspace{-0.8cm}
\centerline{
\scalebox{0.8}{
\includegraphics{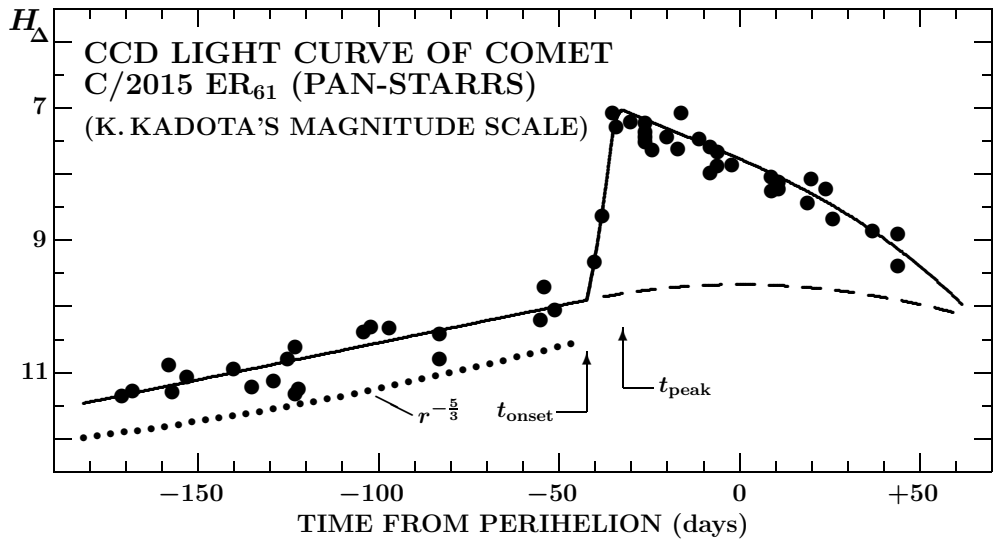}}}
\vspace{-11.25cm}
\caption{Light curve of comet C/2015 ER$_{61}$ based on the CCD observations by
five Japanese observers (the so-called total magnitudes, as published in the {\it
Minor Planet Center\/}'s database), normalized to Kadota's magnitude scale.  The
magnitude $H_\Delta$ is corrected for a geocentric distance using an inverse
square power law, but not for a phase effect.  The outburst is found to have
commenced on 2017 March 28.6 UT or 42.3~days before perihelion, with an estimated
uncertainty of $\pm$1~day.  It appears to have peaked near April 6.6 UT or
33.3~days before perihelion, with the same estimated uncertainty.  The amplitude
amounted to 2.8~mag, that is, the comet brightened by a factor of 13.  The
dashed curve is an extrapolation of the pre-outburst light curve assuming
{\vspace{-0.04cm}}perihelion symmetry.  The dotted curve, shifted~vertically
down by $\sim$0.5~mag for clarity, shows a shallow $r^{-5/3}$ variation with
heliocentric distance $r$.  Only three out of the 54 data points were rejected, all
reported by the same observer who made the comet much too bright by 1~mag or more
on, respectively, March~3, April~22,~and May~28.  The comet's integrated
magnitude was brighter than Kadota's scale suggests; at peak light the comet was
1.2~AU from the Earth.{\vspace{0.5cm}}}
\end{figure*}

This set of uniform observations suggests that the outburst began on 2017
March 28.6 UT (with an estimated uncertainty of $\pm$1 day), the amplitude was
2.8~mag (with an estimated uncertainty of $\pm$0.1~mag), and the peak brightness
was attained on about 2017 April 6.6 UT, or fully 9 days after the event's onset.

\subsection{Light Curve from Visual Observations}
To construct a light curve based on visual observations, I consulted
three independent database websites:\ the {\it International Comet
Quarterly\/}\footnote{See {\tt http://www.icq.eps.harvard.edu/CometMags.html.}},
Crni Vrh Observatory's {\it COBS database\/}\footnote{See {\tt
http://www.cobs.sl/analysis.}}, and the Iberoamerican~\mbox{Astronomical}
League's {\it Comet Observations\/} (LIADA) site\footnote{See {\tt
http://rastreadoresdecometas.wordpress.com.}}, limiting the period of time
to an interval from the beginning of January (130~days before perihelion)
to May 9, 2017 (perihelion).  From the three largely overlapping sources I
collected a total of some 160 visual observations from the critical period,
which were reported by 26~observers using 43~different instruments.  For
analysis I employed only the datasets that included at least a few
observations covering periods of time longer than several days.  It
turned out that this condition was not satisfied by 16~observers using
19~instruments.  For the remaining sets by 14~observers using 24~instruments
I examined the light curve with the aim to derive the magnitude corrections
to bring the data to a common photometric scale; I chose the binocular
observations by J.~J.~Gonz\'alez as a reference set after I found that his
magnitudes obtained with the 10$\times$50 binoculars and the 25$\times$100
binoculars provided a consistent set of data over a period of more than two
months.  I found that no constant correction factors could be derived for four
remaining observers (each~with one instrument) to link their magnitudes
with the rest~of the data.  I was thus left with 112 brightness estimates
by 11~observers with 20~instruments to employ in constructing the visual
light curve displayed in Figure~2; the corrections to Gonz\'alez's scale
were determined by a graphical method, as described at the start of
Section~2.

\begin{figure}[hb]
\vspace{-4.6cm}
\hspace{1.67cm}
\centerline{
\scalebox{0.8}{
\includegraphics{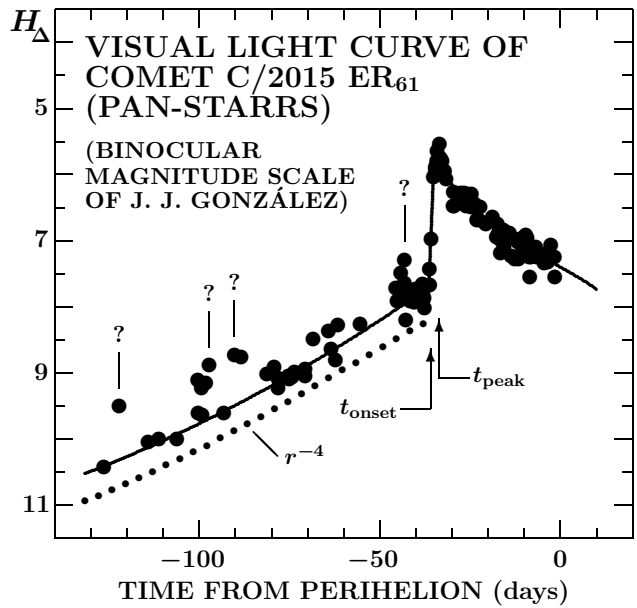}}}
\vspace{-11.55cm} 
\caption{Preperihelion light curve of comet C/2015 ER$_{61}$ based on 112 visual
magnitude estimates, reported by 11~observers (with 20~instruments) and
normalized to the scale of J.\,J.\,Gonz\'alez's binocular observations.  The
outburst is found to have begun most probably on 2017 April 3.9$\,\pm\,$0.1~ UT
or 36.0~days before perihelion and to have peaked only 2.6~days later,
33.4~days before perihelion (with an uncertainty of about $\pm$1~day).  Compared
with the results in Figure~1, the outburst is now determined to have started
more than 6~days later, the time between the onset and the peak is about
6.5~days shorter, and the amplitude amounts to 2.1~mag, 0.7~mag shallower.
There appear to be up to four minor flare-ups between the beginning of 2017 and
the end of March (indicated by the question marks), although these are likely
to be due to errors of observation (see the text).  The~pre-outburst brightness
{\vspace{-0.05cm}}varies with heliocentric distance in this case more steeply,
as $r^{-4}$.  For additional information, consult the caption to
Figure~1.{\vspace{0.02cm}}}
\end{figure}
\begin{figure*}[t]
\vspace{-4.25cm}
\hspace{-0.2cm}
\centerline{
\scalebox{0.765}{
\includegraphics{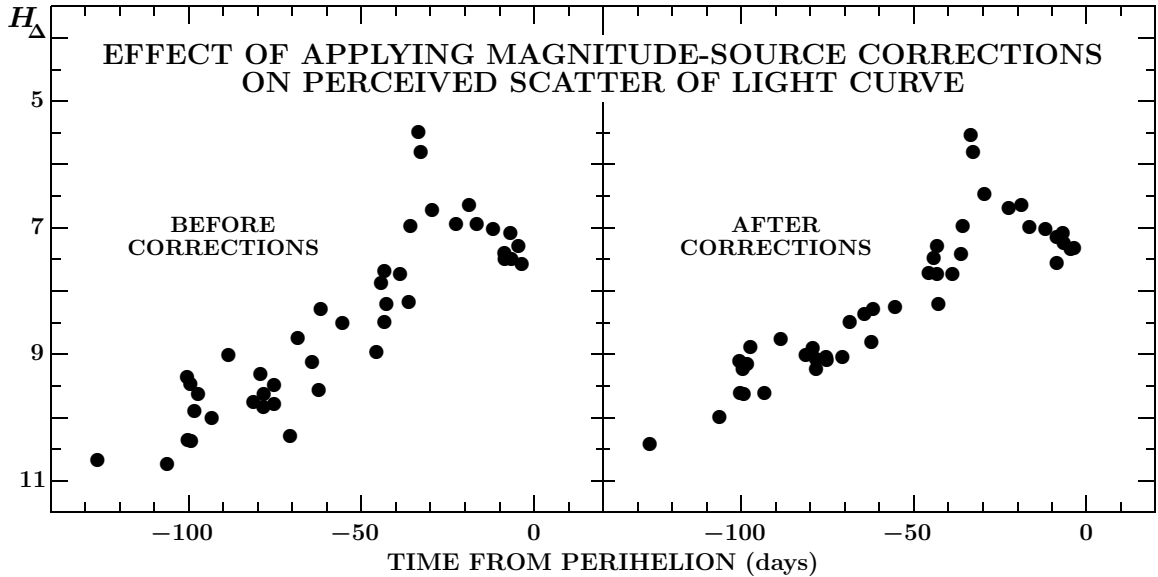}}}
\vspace{-10.84cm}
\caption{Comparison of a light curve before (left) and after (right) applying
magnitude-source corrections to the data reported by six observers:\ P.\ Camilleri,
40.6-cm reflector; M.\ Goiato, 22-cm reflector; J.\ J.\ Gonz\'alez, 5-cm and 10-cm
binoculars; C.\ Hergenrother, 5-cm binoculars; T.\ Lehmann, 10.6-cm refractor; and
C.\ Wyatt, 25-cm reflector.  Note the dramatically reduced scatter on the right.
See the text for the correction factors used and captions to Figures~1 and 2 for
additional information.{\vspace{0.5cm}}}
\end{figure*}

To illustrate the effect of introducing the magnitude-source corrections to the
data, I compare in Figure~3 a partial visual light curve before (left) and after
(right) the appropriate corrections have been applied to the magnitudes reported
by six observers (P.~Camilleri,~\mbox{40.6-cm} reflector, \mbox{{\sl corr} =
$-$0.40 mag};~M.~Goiato,~\mbox{22-cm}~\mbox{reflector},
$-$1.25\,mag;\,C.\,Hergenrother,\,5-cm~binoculars,~+0.05\,mag;
T.\,Lehmann,\,10.6-cm\,refractor,\,$-$0.25\,mag;\,and\,C.\,Wyatt,
25-cm reflector, $-$0.75~mag; in addition to Gonz\'alez~with 5-cm and
10-cm binoculars, no correction).  The figure, an example with arbitrarily
selected observers,~demonstrates that the uncorrected light curve exhibits
much greater scatter than the corrected curve.  Fitting the~pre-outburst
segment of the curve with a straight line yields a mean residual of
$\pm$0.48~mag before correction but only $\pm$0.30~mag after correction,
a drop by a factor of 1.6.

Having now a measure of the light curve's scatter, it is possible to address an
issue of the nature of the four apparent minor flare-ups on the pre-outburst
light curve in Figure~2.  The marked highest points lie above the~fit to the base
data by, respectively from the left to the right, 0.83~mag, 0.82~mag, 0.79~mag,
and 0.68~mag,~that is, between 2.2$\sigma$ and 2.8$\sigma$ and therefore still
within the standard 3$\sigma$ limit of observational error.  While the comet's
intrinsic variation cannot be entirely ruled out, it is unlikely.  This conclusion
is particularly significant for the last of the four peaks, which occurs on
March~27.7 UT (or 43.2~days before perihelion), within 24~hr of the onset time
derived from the CCD-based light curve.

The visual light curve differs from the CCD-based~light curve.  M.~Mattiazzo
reported the comet to be of magnitude 8.3 on April~3.80~UT, of equal brightness as
two days earlier, while 9~hr later, on April 4.17 UT Gonz\'alez estimated it at
7.4, 0.7~mag brighter (when accounting for the scale difference{\vspace{-0.05cm}}
between the two observers), implying a brightening rate of 1.9~mag day$^{-1}$.
Thus, the event could not start before April~3.8 UT, over 6~days later than
suggested by the CCD magnitudes.  From the constraints I estimate that the outburst
began on April~3.9$\,\pm\,$0.1~UT, more than 6~days later than suggested by the
CCD magnitudes, and peaked on April~6.5~UT (with an estimated uncertainty of
$\pm$1~day), at a time nearly identical with that derived from the CCD-based
light curve; the amplitude now amounted to 2.1~mag (with an estimated~uncertainty
of $\pm$0.1~mag), 0.7~mag less than in Figure~1.

\subsection{Light Curve from Nuclear Magnitudes}
I consider the disparity between the two results for~the outburst disconcerting
enough to warrant further investigation.  In a previous paper dealing with the
outbursts of comet 168P/Hergenrother I suggested (Sekanina 2010) that, when
carefully analyzed and interpreted, the so-called nuclear CCD magnitudes ---
referring in fact to the brightness of an inner coma, commonly reported by the
observers with astrometric observations of comets to the {\it Minor Planet
Center\/}, and identified in the {\it Minor Planet Circulars\/} by a symbol N to
distinguish them from the total CCD magnitudes (discussed in Section~2.1) --- can
be employed to great advantage as a diagnostic tool for determining the onset
of cometary outbursts.  This is so because outbursts are known to start with
the sudden appearance and steep brightening of an unresolved plume of material
called euphemistically in the comet jargon a "stellar nucleus'' and described by
a nuclear magnitude.

The results of application of this approach to comet C/2015 ER$_{61}$ were short
of expectations in terms~of~the gathered amount of data, but still led to a
resolution of the disparity between the two light curves.  Focused~on the 70-day
period from late February to the end~of~April 2017 (from 80 to 10 days before
perihelion), I searched the {\it Minor Planet Center\/}'s C/2015 ER$_{61}$ database
(cf.\ footnote 3 in Section~2.1) for observatories that reported nuclear magnitudes
both before and after the outburst.  I was disappointed to find only three:\\[-0.3cm]

\mbox{(1) C}ode J22, Tacande Observatory, La Palma,~\mbox{Islas}
 \mbox{Canarias,\,Spain\,(K.\,Hills,\,50-cm\,f/2.9\,astrograph,\,10\,obs.{\hspace{-0.04cm}});}\\[-0.3cm]

\mbox{(2) C}ode C10,\,Observatoire\,de\,Maisoncelles,\,near\,Saint-Martin-du-Boschet,
Seine-et-Marne, France (J.-F.~Sou\-lier, 30-cm f/3.8 reflector, 7 obs.);
and\\[-0.3cm]

\mbox{(3) C}ode A77, Observatoire Chante-Perdrix, Dauban, Alpes-de-Haute-Provence,
France (J.~Nicolas, J.-F.~Sou\-lier et al., 20-cm f/4 Schmidt-Cassegrain,
40-cm f/3 reflector, 41-cm f/3.3 reflector, 7 obs.).\\[-0.3cm]
\begin{figure}[t]
\vspace{-5.45cm}
\hspace{1.7cm}
\centerline{
\scalebox{0.8}{
\includegraphics{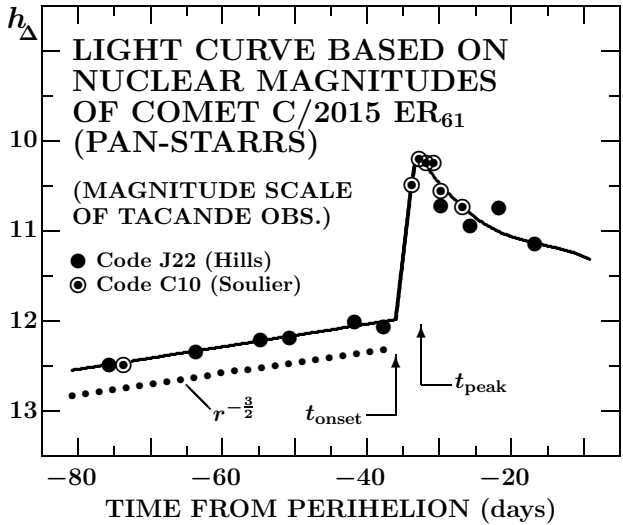}}}
\vspace{-11.6cm}
\caption{Preperihelion light curve of comet C/2015 ER$_{61}$ based on 17 CCD
nuclear magnitudes, as published in the {\it Minor Planet Center\/}'s database.
The plotted magnitudes, $h_\Delta$, were corrected for a geocentric distance
using an inverse square power law (but not for a phase effect) and normalized
to the magnitude scale of the Tacande Observatory (code J22). Note that the two
Tacande Observatory data points on March~29 and April~2 (41.7 and 37.7~days
before perihelion) are not elevated, thus supporting the onset time of the
outburst as derived from the visual light curve (Figure~2), but not from the
CCD-based light curve (Figure~1).  The onset time derived from the visual
light curve is in fact employed in fitting the nuclear magnitudes.  The
{\vspace{-0.04cm}}pre-outburst variations with heliocentric distance follow
a law of $r^{-3/2}$, much more flat than the visual light curve but
comparable to the CCD-based curve.{\vspace{-0.1cm}}}
\end{figure}

\noindent
Comparing the three datasets, it was noticed at once that the nuclear magnitudes
reported by code A77 referred to a much smaller fraction of the nuclear
condensation than the other data, being 2~mag fainter than the C10 data
and 2.5~mag fainter than the J22 data in late February and by up to nearly
2~mag fainter than either of the two sets around the time of the outburst.
Consequently, it was prudent to exclude the A77 data from the
set.\footnote{An additional problem with the code A77 data is the~use~of
up to three different telescopes; because of the overly succinct~format in
which the {\it Minor Planet Center\/} requires observers to present their
observations, it is unclear which of the three telescopes was used for which
of the reported nuclear-magnitude observations, thus introducing potentially
some undesirable instrumental effects.}

Unlike the A77 data, the nuclear magnitudes from the codes J22 and C10 could
readily be combined into one set after a minor scale correction has been
applied.  The resulting light curve, plotted in Figure~4, provides for the
outburst a scenario that in terms of the onset time accommodates the result
from the visual light curve but not from the CCD-based light curve:\ two
critical observations, on March~29.2 UT and April~2.2 UT, show that the
brightness of the nuclear condensation was not yet elevated.  The onset time
derived from the visual light curve is employed in Figure~4 to show that the
two light curves are compatible.  The nuclear magnitudes suggest that the
outburst peaked on about April~7.5~UT (with an estimated uncertainty of
$\pm$1~day), in fair agreement with the results from the visual and CCD-based
light curves.  The amplitude is now a little lower, 1.8$\,\pm\,$0.1~mag,
perhaps an effect of exclusion of the contributions from the outside of
the nuclear condensation.

\begin{figure}[t]
\vspace{0.18cm}
\hspace{-0.17cm}
\centerline{
\scalebox{0.57}{
\includegraphics{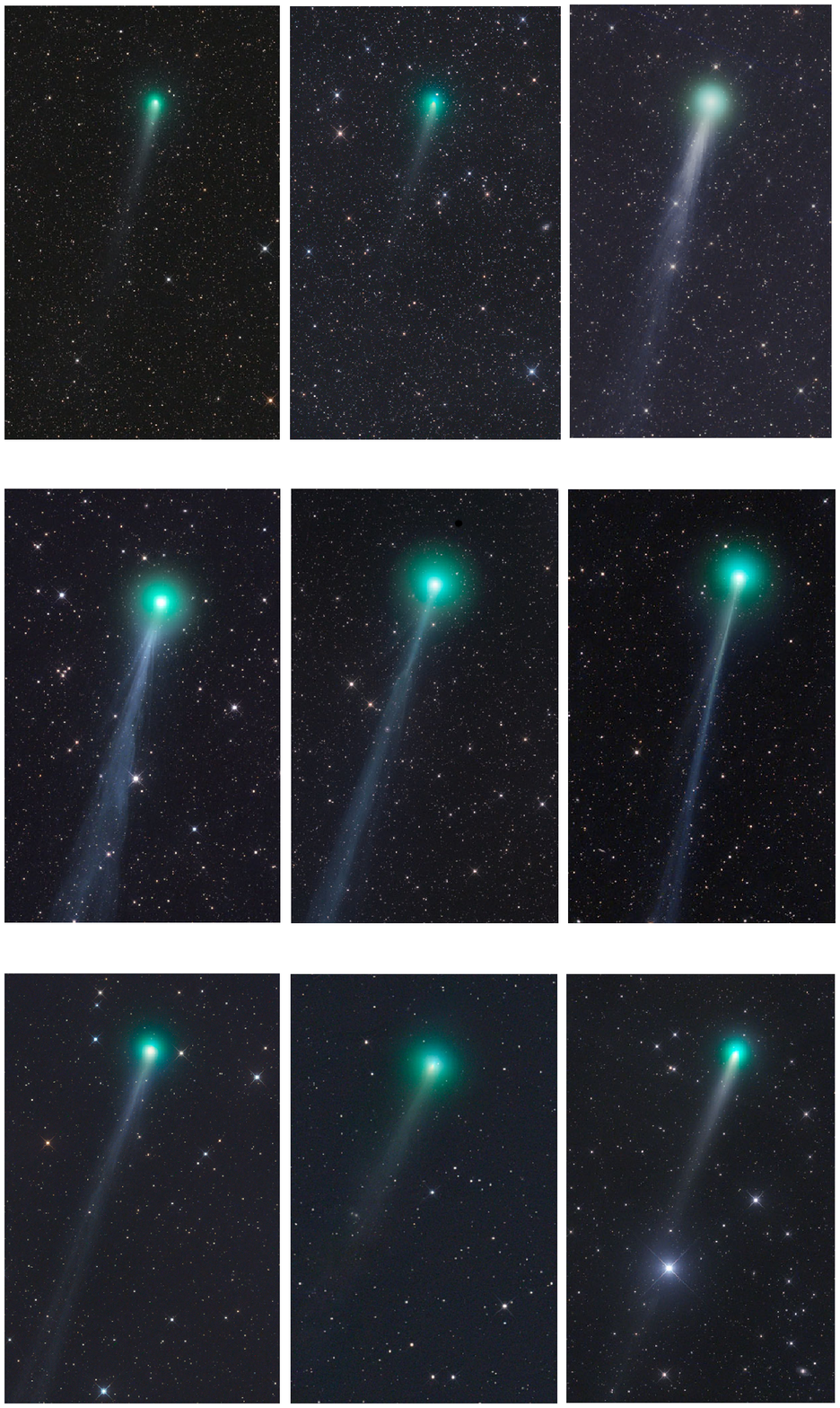}}}

\vspace{-9.9cm} 
\hspace{-0.27cm}
\centerline{
\scalebox{0.5}{
\includegraphics{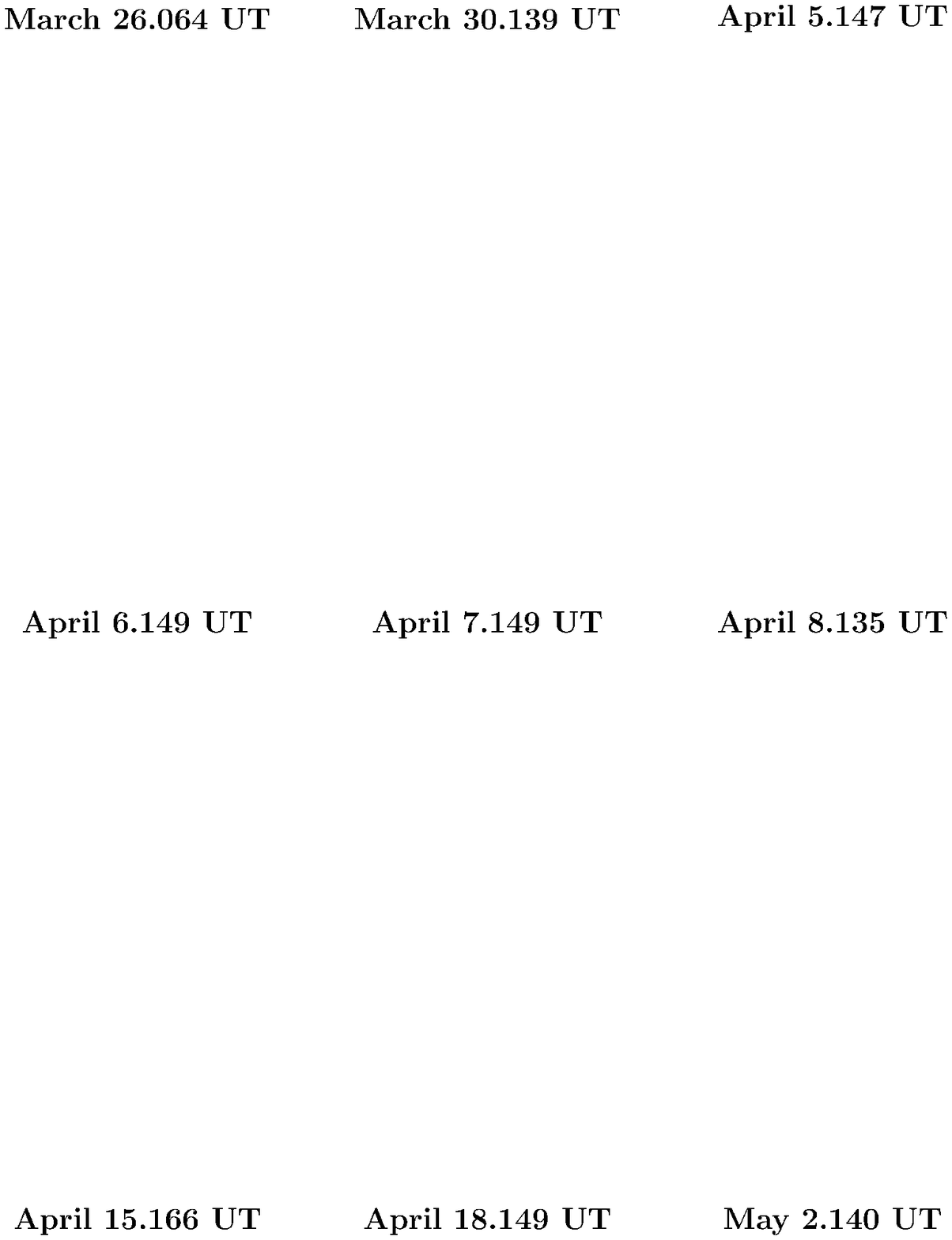}}}
\vspace{0cm}
\caption{LRGB images of comet C/2015 ER$_{61}$ taken by G.\ Rhemann of Austria
between March 26 and May 2, 2017 with a 30-cm f/3.6 astrograph at Farm Tivoli,
Namibia.  The total (L+R+G+B) exposure times were, chronologically:\ 24, 39,
27, 27, 27, 27, 39, 20, and 35~minutes.  Each field extends about 80$^\prime$
along the diagonal.  An approximate orientation has the north to the right and
the east up.  Qualitatively, the comet appears to be the brightest on the images
of April~5--8, fading gradually afterwards.  While the brightness perception is,
to a degree, affected by the uneven exposure times, the comet looks distinctly
fainter on the two images from late March, consistent with the onset time of the
outburst in early April, as derived from the visual light curve and the nuclear
magnitudes.  (Image credit:\ G.\ Rhemann, Astro Systeme Austria.){\vspace{0.5cm}}}
\end{figure}

\begin{figure*}
\vspace{-4.18cm}
\hspace{0cm}
\centerline{
\scalebox{0.73}{
\includegraphics{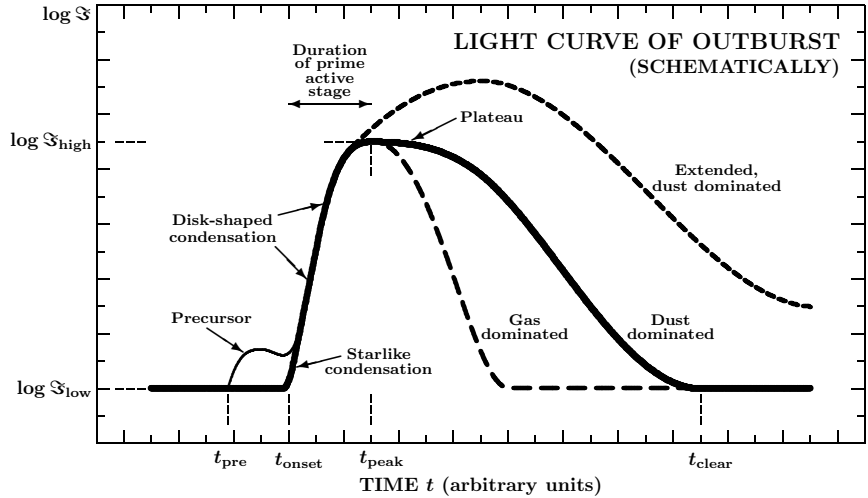}}}
\vspace{-11.1cm}
\caption{Schematic representation of the light curve of outbursts.  The
brightness, $\Im$, is plotted on a logarithmic scale against time, $t$.  Four
categories of events are depicted:\ a generic dust-dominated outburst (solid
curve), an extended dust-dominated outburst (short-dashed curve), a gas-dominated
outburst (long-dashed curve), and a precursor eruption.  The first three types
of events begin at the onset time, $t_{\rm onset}$, when the brightness is
$\Im_{\rm low}$, while the inception of a precursor eruption and a composite
explosion takes place at time $t_{\rm pre}$.  A giant explosion can be described
by scaling the curve for the generic dust-dominated outburst.  The precipitous
rise in brightness, which includes the appearance of a star-like condensation,
is terminated at time $t_{\rm peak}$, the end of the prime active phase, when
$\Im$ reaches a maximum, $\Im_{\rm high}$.  The brightness then begins to
subside slowly along a plateau in a dust-dominated outburst, until it drops
to $\Im_{\rm low}$ at time $t_{\rm clear}$.  By this time, all material
ejected during the outburst has left the volume of the coma.  For an extended
dust-dominated outburst, the coma continues to brighten after $t_{\rm peak}$
and its brightness remains elevated after $t_{\rm clear}$ because of, e.g., an
increasing cross-sectional area of continuously fragmenting dust particles.  In
a gas-dominated outburst, the brightness subsides much more rapidly, reaching
$\Im_{\rm low}$ long before $t_{\rm clear}$.  The symbol $\Im$ refers normally
to the comet's total brightness, but it could also apply to the brightness of
the nuclear condensation.{\vspace{0.5cm}}}
\end{figure*}

\subsection{Evidence from Large-Scale Imaging and Dust-Content Data}
As an additional, though rather qualitative, test~of~the outburst timing, Figure~5
displays nine images of the comet taken by G.\ Rhemann with a 30-cm f/3.6
astrograph at Farm Tivoli in Namibia between March~26 and May~2, 2017.\footnote{See
{\tt http://www.astrostudio.at/2\_Bright Comets.php}.}  Somewhat uneven exposure
times notwithstanding, the comet clearly appears to be much fainter in the first
two images, from late March, than in the following ones, suggesting that the
outburst did not begin before March 30.2~UT, in conformity with the results
derived from the visual light curve and nuclear magnitudes, but contrary to
what was found from the CCD-based light curve.  I see no need for any correction
to the onset time as presented in Section~2.2.

Finally, a set of measurements of {\it Af}$\rho$, a proxy of the dust-production
rate (A'Hearn et al.\ 1995), is also consistent with the same outburst timing.
For an aperture of 10\,000~km, a Spanish~website\footnote{See {\tt
http://www.astrosurf.com/cometas-obs.}} lists the {\it Af}$\rho$ values measured
by numerous observers to average $\sim$600~cm over a four-month period ending
March 31, 2017, but to amount to $\sim$2000~cm on three nights between April~6
and 13, equivalent crudely to the dust-production rates of 0.6 and 2$\times\:
\!\!$10$^6$\,g/s, respectively.  Comparison with~some preliminary water-production
rates suggests that the dust-to-gas production rate ratio for this comet is low,
in line with the object's bulb-like (or onion-like; from the German word {\it
Zwiebelgew\"{a}chs\/}; e.g., Bobrovnikoff 1927, 1928) appearance (Section~1).

\section{Types and Parameters of Cometary Outbursts}
In my previous paper on the subject (Sekanina 2010), I defined an outburst of
a comet as its ``sudden, prominent, and unexpected brightening, caused by an
abrupt, short-term injection of massive amounts of material'' from the nucleus
into the atmosphere.  In Section~2.3 I already mention that, observationally,
outbursts begin with the sudden appearance and precipitous brightening of a
``stellar nucleus.''  This feature develops into a rapidly expanding disk with
a progressively decreasing surface brightness, until it eventually disappears.

Outbursts appear to be activated by gases liberated from a reservoir of a highly
volatile material stored in the nucleus that becomes heated up and/or pressurized.
The products observed in the optical wavelegths are (i)~the escaping gases that
radiate in that spectral range and (ii)~solid material, or dust, that, dragged
from the nucleus by the outflowing gases, scatters sunlight.  Depending on which
of the two components prevails, one distinguishes between {\it dust-dominated\/}
and {\it gas-dominated\/} outbursts.  They have some common features but differ
in other respects, as depicted schematically in Figure~6.  An important common
feature is an {\it active phase\/}:\ the activity of the outburst source on the
nucleus begins at the time of onset and terminates, in a basic timeline, at the
time of peak light, i.e., the duration of the active phase is defined by the
interval between the onset and the peak.  On the other hand, the differences
between the two types of outburst are exemplified by a {\it plateau\/} ---
characterizing {\it only\/} the dust-dominated outbursts --- which extends from
the point of peak light on for a limited period of time and along which the
comet's brightness subsides only gradually.  By contrast, the peak on the light
curve of the gas-dominated outbursts is followed by a steep decline in brightness,
whose rate almost equals the rate of the initial brightening.  This is so because
the brightness variations in the gas-dominated outbursts are determined by the
fairly short dissociation and ionization lifetimes of the radiating molecules in
the coma (usually a day or so near 1~AU from the Sun), coupled with relatively
{\vspace{-0.04cm}}high gas velocities ($\sim$1~km~s$^{-1}$).  On the other hand,
because of lower velocities of solid particles (in fact, dramatically lower
velocities for heavier grains), the residence times of dust in the coma are
substantially longer than the residence times of radiating molecules, hence a
post-peak plateau in the dust-dominated outbursts.

The timeline of an outburst can be more complicated; if it is preceded by one or
more {\it precursor eruptions\/}, the event becomes a {\it composite explosion\/}.
Also, dust particles might fragment in the atmosphere, expanding their total
cross-sectional area and increasing the comet's brightness over a longer period
of time; such a scenario is referred to as an {\it extended dust-dominated\/}
outburst.

The sources of outbursts have typically a fairly limited extent on the scale
of nuclear dimensions, so most outbursts can be classified as local or, at the
most, regional episodes.  Exceptionally, a major part of the nucleus may get
involved, so such events are of a global extent on the scale of nuclear
dimensions, with potentially severe implications for the comet's future evolution;
they are referred to here as {\it giant explosions\/}.  The total mass released
from a comet in a giant explosion amounts to or exceeds 10$^{13}$\,grams, orders
of magnitude more than in a typical outburst.

Comparing the light curves of the various types of outburst, Figure~6 has been
adapted from Sekanina (2010).  The presented scenario assumes that a background
level of the comet's activity before, during, and after the event is constant.
The description of the active phase requires four basic data points:\ time
$t_{\rm onset}$ and~\mbox{magnitude} ${\cal H}_{\rm onset}$ for the onset and
time $t_{\rm peak}$ and magnitude ${\cal H}_{\rm peak}$ for the peak light of
the outburst, with an unknown time of peak injection rate preceding $t_{\rm peak}$.
It is convenient to reckon time from the perihelion passage, $t_\pi$, by
introducing
\begin{equation}
\tau_{\rm point} = t_{\rm point} \!-\! t_\pi, \;\;\;\;\;
 {\rm point} = {\rm onset, peak}.
\end{equation}
The time interval, interpreted as the duration of an active phase of the
outburst, is a nominal parameter referred to as a {\it rise time\/},
$\Delta \tau$,
\begin{equation}
\Delta \tau = \tau_{\rm peak} \!-\! \tau_{\rm onset} = t_{\rm peak} \!-\!
 t_{\rm onset} > 0.
\end{equation}
The other nominal parameter is an {\it amplitude\/}, ${\cal A}$, of the
outburst, defined as
\begin{equation}
{\cal A} = -({\cal H}_{\rm peak} \!-\! {\cal H}_{\rm onset}) > 0.
\end{equation}
Since $\Delta \tau$ typically equals a few days, the geocentric~distance
barely changes over $\Delta \tau$, unless the comet is rather close to the
Earth.~Accordingly, in a general case it~hard\-ly makes any difference whether
the magnitudes are apparent or corrected for the geocentric distance.  In
the latter instance \mbox{${\cal H}_{\rm onset} \!=\!  (H_\Delta)_{\rm onset}$}
and \mbox{${\cal H}_{\rm peak}\!=\! (H_\Delta)_{\rm peak}$} for a CCD-based
{\vspace{-0.03cm}}(Figure~1) and visual (Figure~2) light curve, but
\mbox{${\cal H}_{\rm onset} \!=\!  (h_\Delta)_{\rm onset}$} and \mbox{${\cal
H}_{\rm peak} \!=\!  (h_\Delta)_{\rm peak}$} for a light curve based on the
nuclear magnitudes (Figure~4).  In the following, I consistently use the
magnitudes corrected for the geocentric distance.

The two nominal parameters can be combined into an {\it average flare-up
rate\/}, $\Lambda_0$, to describe the rate of increase in brightness in the
early stage of an outburst,
\begin{equation}
\Lambda_0 = \frac{\cal A}{\Delta \tau}.
\end{equation}
The rate of fading after the outburst has peaked distinguishes dust-dominated
outbursts from gas-dominated outbursts, as already remarked above.

The problem with this definition of the flare-up rate is the magnitude as
a logarithmic quantity.  In particular, if the comet is very faint at the
onset of the outburst, its amplitude may become exceedingly high even if the
amount of material released during the event is not very large.  For this reason,
it is more meaningful to express the brightness in flux units, namely,
\begin{equation}
{\cal F}_{\rm point} = f_0 \, 10^{-0.4 {\cal H}_{\rm point}}\!, \;\;\;\;\;
 {\rm point} = {\rm onset, peak},
\end{equation}
which allows one to define a {\it flux rise\/} between the onset and peak
brightness of the outburst by
\begin{equation}
\Delta {\cal F} = {\cal F}_{\rm peak} \!-\! {\cal F}_{\rm onset} = f_0 \,
 10^{-0.4 {\cal H}_{\rm peak}} \! \left( 1 \!-\! 10^{-0.4 {\cal A}} \right),
\end{equation}
where ${\cal F}_{\rm point}$ is technically an illuminance from the~comet (at
{\vspace{-0.04cm}}1~AU from the Earth) when the units of the constant $f_0$
are, e.g., \mbox{lux = lumen m$^{-2}$} or \mbox{phot = lumen cm$^{-2}$}.
However, ${\cal F}_{\rm point}$ could instead represent the comet's irradiance,
that is, a radiant flux received by a surface per unit area when $f_0$ is in
units such as, e.g., W\,m$^{-2}$ or photons~cm$^{-2}$~s$^{-1}$.  If the
magnitudes ${\cal H}_{\rm onset}$, ${\cal H}_{\rm peak}$ in Equations~(13) and (14)
refer to the top of the Earth's atmosphere (i.e., an unattenuated flux) and
are visual magnitudes (i.e., a flux integrated over the Johnson-Cousins
standard $V$ passband, with a central wavelength of 0.55~$\mu$m), then
\mbox{$f_0 = 0.88 \times\! 10^6$\,photons cm$^{-2}$ s$^{-1}$}; if, instead,
the data are red magnitudes (i.e., an unattenuated flux integrated over
the Johnson-Cousins standard $R$ passband {\vspace{-0.01cm}}with a central
wavelength of 0.64~$\mu$m), then \mbox{$f_0 = 1.07 \times \!  10^6$\,photons
cm$^{-2}$ s$^{-1}$} (Bessell 1979).\footnote{See also {\tt
http:/\hspace{-2pt}/www.cfa.harvard.edu/$\sim$dfabricant/huchra/ ay145/mags.html}.}
Since both visual and CCD-based magnitudes are dealt with in this paper (the
transmission curves of the CCD sensors differing from the standard passbands),
{\vspace{-0.04cm}}I will employ below \mbox{$f_0 = 10^6$\,photons cm$^{-2}$
s$^{-1}$} as a rounded-off orientation value.

When the amplitude ${\cal A}$ is very large, the limiting value of the flux rise,
\begin{equation}
\lim_{{\cal A} \rightarrow \infty} \!\! \Delta {\cal F} = \Delta {\cal F}_\infty
 = f_0 10^{-0.4 {\cal H}_{\rm peak}} \!,
\end{equation}
depends~only~on~the~\mbox{magnitude}~of~the~\mbox{outburst}~peak.~Ac\-cording to
{\vspace{-0.03cm}}Equation~(14),~\mbox{$0.9 \:\!  \Delta {\cal F}_\infty \!<\!
\Delta {\cal F} \!<\!  \Delta {\cal F}_\infty$}~when \mbox{${\cal A} \!>\!
2\frac{1}{2}$\,mag} and \mbox{$0.99\:\!\Delta {\cal F}_\infty\!<\!\Delta {\cal
F} \!<\! \Delta {\cal F}_\infty$}~when~\mbox{${\cal A} \!>\! 5$\,mag}.  This
argument justifies the approximate determination of a flux rise for some outbursts
and all giant explosions, for which the magnitude at the onset (and therefore~the
amplitude) is unavailable (e.g., when the comet is~discovered during an outburst,
such as 17P in 1892) but for which there exist grounds to believe that the comet
brightened substantially during the event.  In addition, this exercise suggests
that by itself the amplitude is not a critical outburst parameter.

Equation (14) replaces (11) as a measure of the increase in brightness caused by
the outburst.  Similarly, the rate of brightening $\Lambda_0$ from Equation~(12)
is replaced with an {\it average rate of flux rise\/}, 
\begin{equation}
\Lambda = \left \langle \! \frac{\partial {\cal F}}{\partial t} \!\!\: \right
 \rangle = \frac{\Delta {\cal F}}{\Delta \tau},
\end{equation}
where $\Delta {\cal F}$ comes from Equation~(14) and $\Delta \tau${\vspace{-0.04cm}}
from Equation~(10); if the rate $\Lambda$ is expressed in \mbox{photons cm$^{-2}$
s$^{-2}$}, $\Delta \tau$ must be in seconds rather than days.  Note that unlike
$\Delta {\cal F}$, $\Lambda$ cannot be determined (not even approximately) when
for some reason there is no information on the outburst inception.

If it should be desirable to normalize the flux-rise rate from Equation~(16) to
the comet's irradiance at the time of onset of the outburst, one can introduce
an {\it average relative flux-rise rate\/}, $\lambda$,
\begin{equation}
\lambda = \frac{1}{{\cal F}_{\rm onset}} \left \langle \! \frac{\partial
 {\cal F}}{\partial t} \!\!\: \right \rangle = \frac{10^{0.4 {\cal A}}
 \!-\! 1}{\Delta \tau},
\end{equation}
which, as expected, depends on both ${\cal A}$ and $\Delta \tau$; its dimension
is day$^{-1}$ or s$^{-1}$.  The new parameters $\Delta {\cal F}$, $\Lambda$, and
$\lambda$ are now ready to be compared to the nominal parameters ${\cal A}$ and
$\Lambda_0$.

\section{Comparison of Outbursts in C/2015 ER$_{61}$ and Other Comets}
%
The nominal as well as newly introduced~\mbox{parameters} are next employed to
compare the outburst experienced by C/2015~ER$_{61}$ with those that other,
selected comets were subjected to, including the giant explosions that
17P/Holmes and 1P/Halley were observed to undergo; to investigate the range
and distribution of the parametric values; and to illustrate the major
differences between the nominal and flux-based parameters.

The results are summarized in Table 4, which is mostly self-explanatory.  The
data type in column 2 refers to the segment of the light curve that affects the
outburst parameters, not necessarily to the whole curve:\ {\sf vis} stands for
visual observations, {\scriptsize CCD} for such (usually unfiltered) observations;
{\sf comb} for combined visual and photographic or CCD observations; and {\sf
nucl} for (CCD) nuclear magnitudes.  As already noted, all employed observations
were always first corrected, to the extent possible, for personal and instrumental
effects, normalized to a common magnitude scale, and corrected to 1~AU from the
Earth before a light curve was plotted, analyzed, and interpreted.  The
uncertainties in the normalized brightness data determine the errors of the
parameters in the table, which generally range from a fraction of 1~day to
a couple of days in $\Delta \tau$; 0.1~mag to a few tenths of magnitude in
${\cal A}$; the relative errors of $\Delta {\cal F}$, $\Lambda$, and $\lambda$
are typically on the order of a few percent to a few tens of percent.

To cover a wide range of objects and explosive events, I compare C/2015~ER$_{61}$
with objects subjected to both dust-dominated and gas-dominated events, including
a variety of ordinary outbursts (both at the same apparition and at different
apparitions for periodic comets) and all giant explosions that I am aware of;
one of them is interpreted in two different ways, both as a single event and a
composite explosion that consists of a precursor burst and a main eruption.

It is noted that the nominal parameters of ordinary outbursts and giant explosions
overlap and thus are not diagnostic for discriminating between the two categories;
in particular, a giant explosion is not identified by an enormous amplitude.  On
the other hand, two of the newly introduced parameters, the flux rise, $\Delta
{\cal F}$, and the average flux-rise rate, $\Lambda$, separate the two categories
of explosions quite distinctly.  The first of them largely reflects the
normalized magnitude at peak brightness, ${\cal H}_{\rm peak}$, so that when
experiencing a giant explosion, the comet is intrinsically much brighter than
when it is subjected to an outburst.  The second condition is more subtle:
when in a giant explosion, the comet brightens at a faster average pace than
in an outburst.

Of the listed comets, the two successive 1973 outbursts of 41P were closest to
the giant explosions in terms of ${\cal H}_{\rm peak}$; however, they were
typical gas-dominated events, with the mass of dust released by 41P orders
of magnitude lower than the amounts lost by 17P or 1P.

The superiority of the parameters $\Delta {\cal F}$ and $\Lambda$ over the
nominal ones, $\cal A$ and $\Lambda_0$, is also apparent from comparison
of the four outbursts of comet C/2001~A2 (Sekanina et al.\ 2002).  While
the amplitude suggests that Event~I was by far the most prominent among the
four, the values of $\Delta {\cal F}$ show that, together with Event~IV,
Event~I was in fact the least powerful and that the strongest one (by a factor of
nearly 4 relative to Event~I) was Event~III, which may have incorporated
precursors.  Also in line with its respectable magnitude, this event accompanied
the separation from the primary nucleus B of three companion nuclei, fully one
half of the total of six detected.

Quite exceptional was the outburst of comet 73P in 1995, which related to a
fragmentation of the parent nucleus resulting in the birth of the principal
fragment C and a persistent companion fragment B.  The rise time of 36 days
encompassing the perihelion time, represents a record in Table~4.  The event
probably consisted of a number of successive flare-ups that gradually built
the amplitude up to 5~mag.  The post-peak evolution of the light curve was
also peculiar.  It first dropped at a moderate rate, only to develop into a
second outburst (not tabulated here) of a much smaller amplitude, but of
considerable plateau that extended over some 100~days.

\begin{table*}[t]
\vspace{-4.18cm}
\hspace{-0.53cm}
\centerline{
\scalebox{1}{
\includegraphics{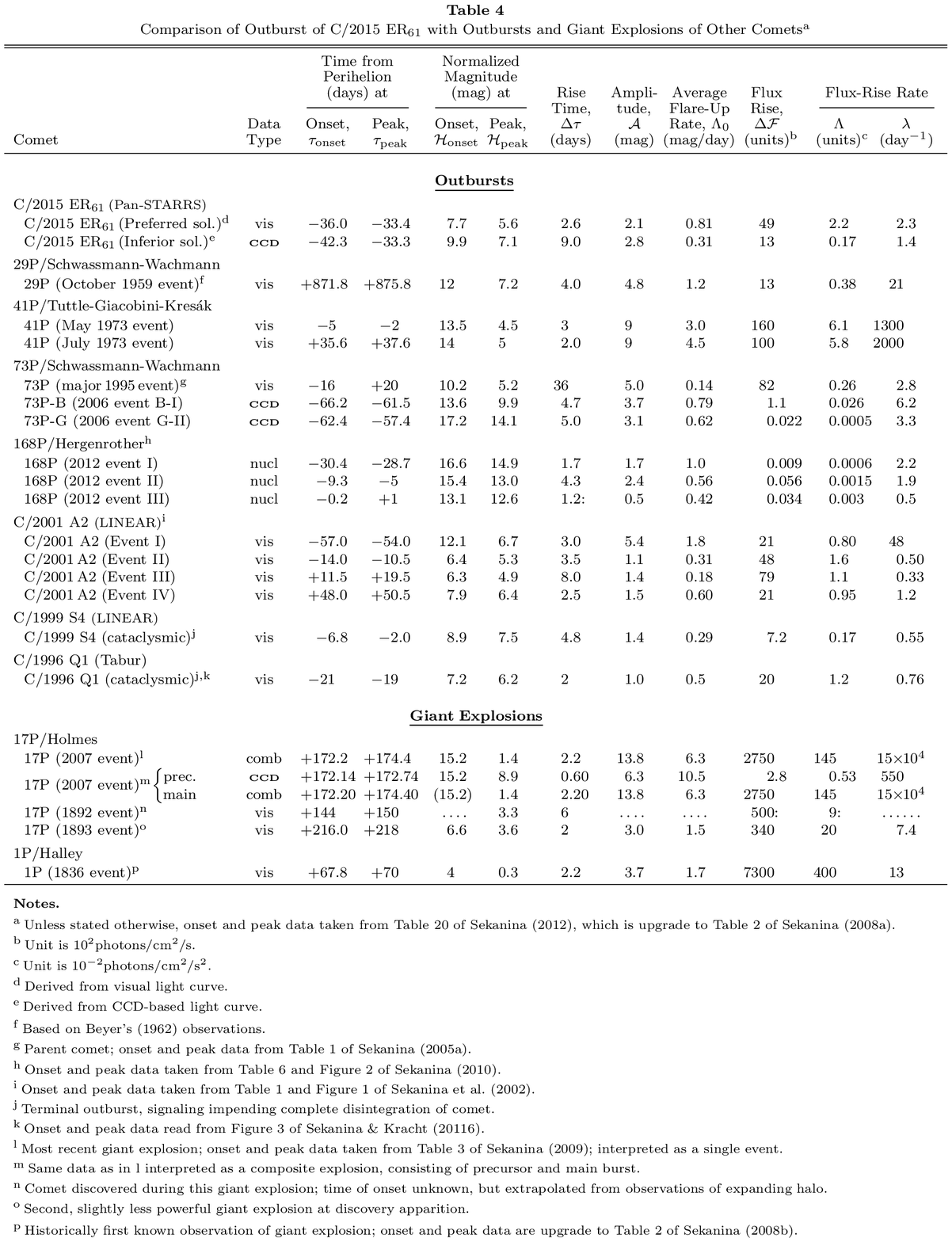}}}
\end{table*}

Of a similar amplitude as the extended explosion of 73P was an October 1959
outburst~of 29P/Schwassmann-Wachmann, an object that is famous for its
propensity to flaring up.  This event's timeline was described in great
detail by Beyer (1962), who monitored the comet for three months by both measuring
increasing coma dimensions and providing a visual light curve.\footnote{This
outburst was one of the most prominent on record (peak apparent visual magnitude
of 10.7); competitive events reported in extensive recent investigations (e.g.,
Trigo-Rodr\'{\i}guez et al.\ 2010; Hosek et al.\ 2013; Miles et al.\ 2016) are
very rare.{\vspace{-0.1cm}}}~I extrapolate the measured coma-expansion rate to
determine the time of outburst onset to $\pm$0.2~day.  I estimate the normalized
magnitude at onset at \mbox{${\cal H}_{\rm onset} \!=\! 12 \:\!\!\pm\:\!\! 0.5$}
(equivalent to apparent magnitude 15.5) from the following information on the
comet:\ (i)~its nondetection by Beyer with a 26-cm refractor on September~27
and 30; (ii)~its nondetection on patrol plates at the Sonneberg Observatory
taken on September~30 (Beyer 1962); (iii)~its photographic detection by Roemer
(1959) with a 102-cm reflector on September~25; and (iv)~its apparent magnitude
at the quiescent phase averages about 16 (e.g., Trigo-Rodr\'{\i}guez et al.\
2010).  Compared to the 1995 event of 73P, the rise time is now one order of
magnitude shorter, although the two events are alike in displaying an extremely
long post-peak plateau.

Contrary to these two instances and the giant explosions, most cometary outbursts
--- including, among others, those of C/2015~ER$_{61}$, 41P, and C/2001~A2 ---
appear to be fairly short-lived, at least in terms of the total brightness, the
light curve exhibiting a sharp peak rather than a plateau, that is, no signature of
dust dominance.  At first sight, this fact may look peculiar considering that the
``stellar nucleus,'' appearing in the early phase of an outburst (Section~2.3),
consists of a cloud of expanding dust.  The explanation has long been known:\
a close monitoring of a relatively modest outburst of 12P/Pons-Brooks on January 1,
1884 by M\"{u}ller (1884a, 1884b) and independently by Vogel (1884) led both
observers to conclude that while the coma brightness was dominated by three
bands (of C$_2$), a continuous spectrum prevailed in the nuclear condensation
and its outskirts.  Hence, unless the amount of dust in the ejecta is
overwhelming, its contribution to total light of the comet drops with time too
rapidly to exhibit a plateau.

A special place among cometary explosions has the~category of {\it terminal
outbursts\/}, represented in Table~4 by comets C/1996~Q1 (Tabur) and C/1999~S4
(LINEAR).  Unlike other outbursts, this type of flaring up is of cataclysmic
nature, as it signals an imminent, complete disintegration of the comet in the
course of several days to, at most, a few weeks following the outburst.  Both
tabulated events have a fairly modest amplitude, less than 1.5~mag.  C/1996~Q1
was genetically related to C/1988~A1 (Liller) and C/2015~F3 (SWAN), making up
a group of long-period comets with nearly identical orbits; of these, C/1988~A1
was the main mass, while C/1996~Q1 and C/2015~F3 were its companions.  The
group's history was recently investigated by Sekanina \& Kracht (2016), where
the reader can find the light curve of C/1996~Q1 from which the onset and
peak data were read.

On the other hand, C/1999 S4 is not known to be the member of a comet group
(or a pair), although it was suggested by Sekanina (2000) that this object
could have been a trailing fragment of a more massive comet moving in a nearly
identical orbit, whose arrival to perihelion, perhaps many centuries earlier,
was missed.

The very low values of $\Delta {\cal F}$ and $\Lambda$ for comet 168P are
a product of, in part, the use of the nuclear (rather than total) magnitudes
and, in part, the intrinsic faintness of the object.  In fact, according to
the COBS database (cf.\ footnote 5 to Section~2.2), the comet's total visual
magnitude at peak brightness was less than 2~mag (or a factor of $\sim$5)
above the nuclear magnitude.
%

By comparison, the investigated comet's outburst\footnote{That is, its
preferred solution; the inferior solution is tabulated to show a high degree
of disparity compared to the preferred solution in all parameters, especially an
order-of-magnitude discrepancy in the flux-rise rate.} is on a par with the
outbursts of C/2001~A2, which --- as noted above --- were associated
with a release from the parent nucleus of several nuclear fragments.  It is
therefore plausible to anticipate a possible temporal correlation between the
outburst and splitting of C/2015~ER$_{61}$, a major objective of this paper.

For a temporal correlation with the companion's separation, it is the timing
of the outburst that is of interest.  Since it is reasonable to assume that
the companion was most likely to break off from the parent comet in the course
of the active phase of the outburst, the statistically probable fragmentation
time can in this scenario be equated with the midtime of the event's active
period, $t_{\rm mid}$ (with a probable uncertainty of one half of the rise
time), that is,
\begin{equation}
t_{\rm mid} \!-\! t_\pi = \left( \tau_{\rm onset} \!+\!  {\textstyle \frac{1}{2}}
 \Delta \tau \right) \pm {\textstyle \frac{1}{2}} \Delta \tau.
\end{equation}
The preferred solution in Table~4 gives for C/2015 ER$_{61}$ \mbox{$t_{\rm
mid} \!-\! t_\pi \!=\! -34.7\:\!\!\pm\:\!\!0.5$ days} or $t_{\rm mid} \!= 2017$
April 5.2 UT.

\section{Birth and Motion of Companion Nucleus}
To describe the heliocentric motion of the companion, or fragment B, I refer
it to the primary nucleus, or fragment A, by modeling the temporal variations
in the offsets ``companion minus primary'', or \mbox{B\,--\,A}, of the
astrometric positions measured in images obtained by the same observer
(or a group of observers) in right ascension and declination.  An inherent
part of this exercise is an assumption that the companion and the primary
had a common parent in the past and that they seperated from each other at
some particular time.  The scenario thus involves a process of fragmentation
of the original single nucleus --- the problem of a split comet.

\subsection{The Offsets \mbox{\rm B\,--\,A}}
When the astrometric positions of both fragments are measured in the same
image, the generation of the \mbox{B\,--\,A} offset is straightforward.
However, especially in recent times the astrometric positions of an object
are derived from a set of short exposures by stacking the best of them to
increase the signal and suppress the noise.  If the sets of optimum images
for the primary and the companion are not identical, the effective times of
their stacked images generally differ.

In this study the selected offset times are always identical with the imaging
times of the companion.  To generate a \mbox{B\,--\,A} offset for any of
these times when the imaging times of the two fragments do not coincide, it
is necessary to convert the astrometric position of the primary nucleus from
a nearby time to the time of the companion's astrometric position by applying
a correction for the primary's heliocentric motion between the two times,
employing a (high-quality) set of orbital elements.  To achieve a meaningful
result, only a primary-nucleus' position at a time that differs from the time
of the companion's position by less than a small fraction of a day should be
used.  If there are multiple choices for a given time of the companion's
position, the statistically best offset is derived by averaging all the
individual offsets based on the primary's positions at times within the given
limit from the time of the companion's position.

To compute the offsets of the companion from the primary of C/2015~ER$_{61}$,
the maximum allowed time difference to be incorporated in the \mbox{B\,--\,A}
offsets was adopted to be $\pm$0.01~day.  As the comet's geocentric motion
in the period of June~11--30 averaged about +100$^{\prime\prime}$\,hr$^{-1}$
in right ascension and +33$^{\prime\prime}$\,hr$^{-1}$ in declination, a crude
estimate for a maximum correction to the primary's positions amount to
$\pm$24$^{\prime\prime}$ in right ascension and $\pm$8$^{\prime\prime}$ in
declination.  Most corrections were of course substantially smaller than these
approximate upper limits.

\begin{table*}[t]
\vspace{-4.2cm}
\hspace{-0.5cm}
\centerline{
\scalebox{1}{
\includegraphics{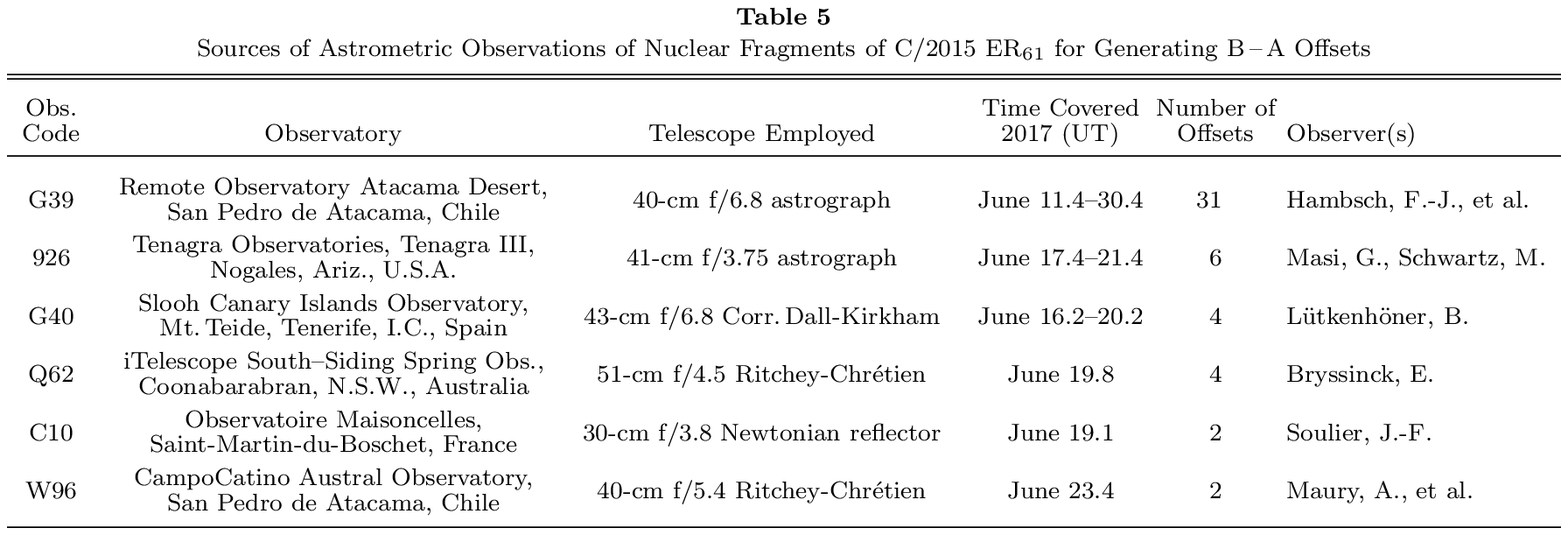}}}
\vspace{-18.6cm}
\end{table*}

\subsection{Fragmentation Model}
Of the 52 astrometric observations of the companion (cf.\ Section~1 and
footnote~2), three could not be used to derive the \mbox{B\,--\,A} offsets
because of the unavailability of the astrometric positions for the primary.
The 49 remaining offsets in right ascension and declination were derived as
explained above from the two fragments' astrometric positions secured by
six groups of observers listed in Table~5.  The appearance of the comet's
nuclear region on one of the images taken at San Pedro de Atacama, Chile,
on June~19 is reproduced in Figure~7.

The 49 \mbox{B\,--\,A} offsets were fitted by a multiparameter model for
split comets (Sekanina 1978, 1982).  Applying an iterative least-squares
differential-correction procedure, the model allows one to solve for up
to five parametric constants:\ the time of fragmentation (i.e., separation
of the two objects), $t_{\rm frg}$; the post-split differential
nongravitational deceleration of the companion relative to the primary
nucleus (driven by anisotropic outgassing from the two objects and assumed
to vary inversely as the square of heliocentric distance), $\gamma$; and the
velocity of separation, $V_{\rm sep}$, of the companion, consisting of the
components in three cardinal directions of the right-handed RTN coordinate
system:\ the radial (away from the Sun) $V_{\rm R}$; transverse $V_{\rm T}$;
and normal $V_{\rm N}$.  Since the nongravitational deceleration varies, in
principle, as the cross-sectional area of the object and inversely as its
mass (i.e., in the sum inversely as the object's size) and since $\gamma$
is a difference between the companion and the (presumably larger and more
massive) primary, the model should always yield \mbox{$\gamma > 0$}.  In
addition, experience with the previously investigated comets suggests that
the separtion velocity $V_{\rm sep}$ is generally quite low,{\vspace{-0.05cm}}
mostly lower than 1~m~s$^{-1}$, and that for a given split comet solutions with
\mbox{$V_{\rm sep} \!>\! 1$ m s$^{-1}$} should be viewed as inferior, suspect,
and not physically meaningful compared to solutions for which
\mbox{$V_{\rm sep} \!<\! 1$ m s$^{-1}$}.

\begin{figure}[b]
\vspace{0.4cm}
\hspace{-0.25cm}
\centerline{
\scalebox{1.61}{
\includegraphics{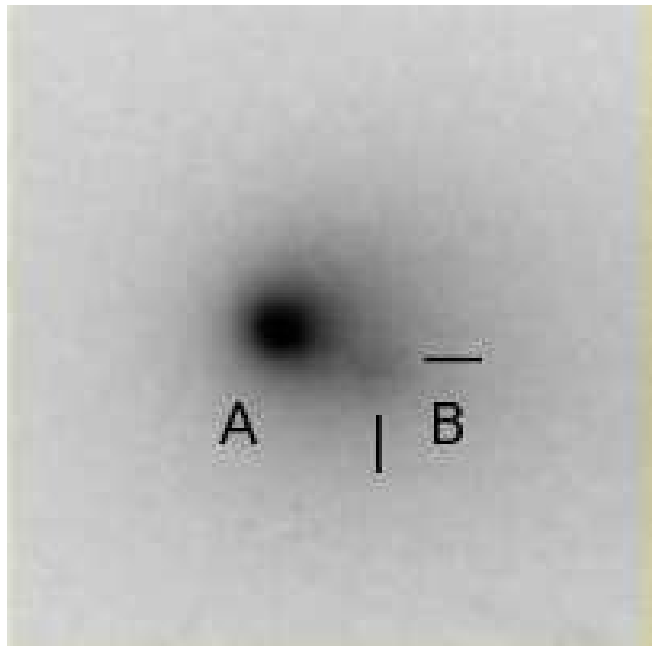}}}
\vspace{0cm}
\caption{Image of the primary nucleus A and companion B of comet C/2015~ER$_{61}$
taken by F.-J.\ Hambsch and E.\ Bryssinck with the 40-cm f/6.8 astrograph
(+\,CCD\,+\,R Bessell filter) of the Remote Observatory Atacama Desert, San Pedro
de Atacama, Chile, on 2017 June 19.41 UT.  The image is about 140$^{\prime\prime}$,
or 152\,000~km at the comet, along the diagonal.  North is up, east to the left.
(Image credit: E.\ Bryssinck, Brixiis Astronomical Observatory; {\tt
http://www.astronomie.be/erik.bryssinck/c2015er61.html}).{\vspace{-0.09cm}}}
\end{figure}

The model allows one to solve for all, or any combination of fewer than, the
five parameters.  The length of the covered arc of the companion's trajectory
and the accuracy of the offsets constrain the number of parameters that can
be solved for. Solutions with a number of parameters greater than this
limit usually fail to converge because they are highly correlated.  This
is particularly true for the pair of the fragmentation time and the radial
component of the separation velocity and for the pair of the deceleration
and the transverse component of the velocity.  On the other hand, one
should always solve for at least two parameters.  It is sometimes a fair
approximation to adopt a zero value for  $V_{\rm R}$, $V_{\rm T}$, and/or
$V_{\rm N}$, even though the velocity's normal component is usually well
determined.  However, only under special circumstances makes it sense to
assume \mbox{$\gamma = 0$} (i.e., when the primary and companion are of
comparable size and activity).  No such approximation obviously applies to
the fragmentation time.  If the splitting of the nucleus is suspected to
coincide with an outburst (or another event), its time could be adopted
for $t_{\rm frg}$; this sort of a coincidence is exactly what I exploit
below.

\begin{table*}
\vspace{-4.11cm}
\hspace{-0.5cm}
\centerline{
\scalebox{1}{
\includegraphics{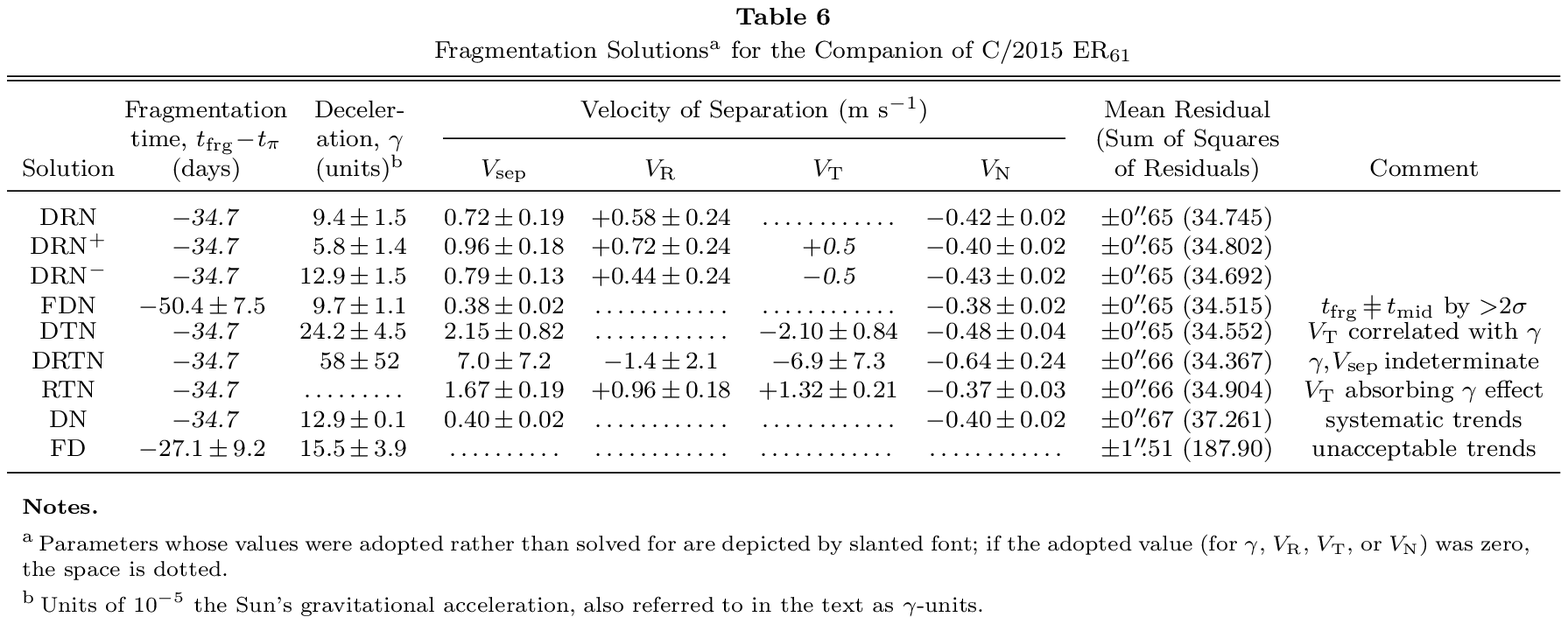}}}
\vspace{-17.7cm}
\end{table*}

The various solutions are referred to below by a letter group in which F
stands for the fragmentation time, D for the deceleration, and R, T, and N
for the three separation velocity components.  Thus, for example, Solution~FD
indicates that the only parameters solved for were $t_{\rm frg}$ and $\gamma$,
while the other parameters were chosen (usually equated with zero except for
$t_{\rm frg}$); Solution DRTN means that the deceleration and all three
separation velocity components were solved for, while for the fragmentation
time I adopted \mbox{$t_{\rm frg} = t_{\rm mid}$} from Equation~(18).  

I searched for all 26 possible two- to five-parameter solutions of the
fragmentation model when applying it to the companion of C/2015~ER$_{61}$;
some, including the most successful, solutions are highlighted in Table 6.  They
are based on 42 consistent offsets with the residual in either coordinate not
exceeding $\pm$1$^{\prime\prime\!\!}$.5.  The remaining seven offsets left
higher residuals and were rejected.  When not solved for, $\gamma$, $V_{\rm
R}$, $V_{\rm T}$, and $V_{\rm N}$ were assigned a value of zero, unless
specified otherwise in the table.  As seen, in all but two tabulated solutions,
$t_{\rm frg}$ was assumed to equal the midtime of the active phase of the
detected outburst (Section~4).  The findings of the model's application are
summarized in the following.

\subsection{Results from Two-Parameter Solutions:\ {\rm FD, FR, FT, FN, DR, DT,
DN, RT, RN,} and {\rm TN}}
None of these solutions is found to be satisfactory.  The least unsatisfactory
is Solution DN, listed in Table~6, which, however, leaves systematic residuals
of up to about 0$^{\prime\prime\!\!}$.5 in right ascension and a less prominent
trend in declination, with a mean residual of $\pm$0$^{\prime\prime\!\!}$.67.
Solutions TN, RN, and RT, based on a dubious assumption of a zero deceleration,
leave systematic trends of $\sim$1$^{\prime\prime}$ or more in the distribution of
residuals and a mean residual of, respectively, $\pm$0$^{\prime\prime\!\!}$.76,
$\pm$0$^{\prime\prime\!\!}$.80, and $\pm$1$^{\prime\prime\!\!}$.28.  The other
solutions with a zero assumed deceleration --- FR, FT, and FN --- fail to
converge.  The remaining three solutions --- FD, DR, and DT --- do not fare
much better:\ the first two leave a very unsatisfactory mean residual of
$\pm$1$^{\prime\prime\!\!}$.51, whereas the last one, leaving a mean residual
of $\pm$1$^{\prime\prime\!\!}$.11, offers a physically meaningless negative
nongravitational deceleration $\gamma$.  Solutions DT, RT, and TN yield
for the transverse component of the separation velocity undesirably high
values in excess of 2~m~s$^{-1}$.

\subsection{Results from Three-Parameter Solutions:\ {\rm FDR, FDT, FDN, FRT,
 FRN, FTN, DRT,\\DRN, DTN,} and {\rm RTN}}
This group includes the best solutions that dominate Table~6, DRN and FDN.
In terms of fitting the measured positional offsets, they are equivalent, but
the timing of the fragmentation event in Solution FDN is determined with
a large mean error and deviates from the midtime of the active phase of the
outburst by only $\sim\!2\sigma$, suggesting a potential relationship between
the two events.  A third solution, DTN, also provides a satisfactory fit, but
the transverse component of the separation velocity is rather poorly determined
because of its high correlation with the deceleration.  Four solutions --- FDR,
FRT, FRN, and FTN --- fail to converge, while other two --- FDT and DRT ---
yield a negative deceleration, again because of its high correlation with
the transverse component of the separation velocity.  The last solution ---
RTN --- offers, as Table~6 illustrates. a separation velocity much higher than
1~m~s$^{-1}$, as its transverse component is apparently forced to absorb
the ignored effect of a deceleration.

\subsection{Results from Four-Parameter and Five-Parameter Solutions:\ {\rm
FDRT, FDRN, FDTN, FRTN,\\DRTN,} and {\rm FDRTN}}
One common problem that these runs are consistently pointing to is too
many unknowns.  Fully five of the six solutions --- FDRT, FDRN, FDTN, FRTN,
and FDRTN --- do not converge, while the sixth, presented in Table~6, shows
both the deceleration and the separation velocity to be indeterminate,
unquestionably again because of high correlations.

\subsection{$\!\!$Final Comments on Fragmentation~of~C/2015\,ER$_{61}$.
 Companion's Longevity}
I find the temporal coincidence between the outburst and the fragmentation
event rather compelling, as converging solutions that include the fragmentation
time as an unknown variable have a tendency to bracket the time of outburst (see
Table~6 for an example).  I do not doubt that Solution DRN is the type of a
solution that should provide the best results in this instance.  Both Table~6
and Sections 5.3--5.5 suggest that it is counterproductive to solve simultaneously
for the deceleration and transverse component of the separation velocity, as
they always are very highly correlated, an option that Solution DRN avoids.
It is also counterproductive to solve for more than three parameters, as the
available database cannot support such elaborate solutions.

There is another observed property of the companion that has not as yet been
examined --- its lifetime against disintegration.  This is the issue behind
my presentation in Table~6 of two additional solutions:\ DRN$^+$, whose
transverse component of the separation velocity is assumed to equal
+0.5~m~s$^{-1}$, and DRN$^-$, for which I adopt instead $-$0.5~m~s$^{-1}$.
The introduction of a change of 0.5~m~s$^{-1}$ is shown to force an opposite
change of about 3.5~units of 10$^{-5}$\,the Sun's gravitational acceleration
(called ``$\gamma$-units'' in the following) in the deceleration:\ a higher
velocity requires a lower deceleration and vice versa.  These parametric
changes have only a negligible and physically irrelevant effect on the
quality of fit, as is seen from the sum of squares of residuals in Table~6.
Besides the deceleration, the other two components of the separation velocity
are also affected, the radial moderately, the normal just slightly, but in
either case by an amount that is comparable to their mean errors.  The effects
of an introduced $V_{\rm T}$ can be expressed numerically thus (all in the
units used in Table~6):
\begin{eqnarray}
\gamma & = & 9.39 - 7.10 \, V_{\rm T}, \nonumber \\
V_{\rm R} & = & +0.578 + 0.283 \, V_{\rm T}, \nonumber \\
V_{\rm N} & = & -0.416 + 0.033 \, V_{\rm T}.
\end{eqnarray}
From the second and third relations one gets
\begin{equation}
V_{\rm sep} = \sqrt{1.081\,V_{\rm T}^2 + 0.300 \, V_{\rm T} + 0.507}.
\end{equation}
It follows that the separation velocity stays below 1~m~s$^{-1}$ as long as
\mbox{$-0.83\:{\rm m}\:{\rm s}^{-1} < V_{\rm T} < +0.55\:{\rm m}\:{\rm s}^{-1}$};
the deceleration is then confined to \mbox{$5.5\;\gamma$-units$\:< \!\gamma\! <
\! 15.3\;\gamma$-units}.  The separation velocity reaches a minimum of
\mbox{0.70$\:$m$\:$s$^{-1}$} when \mbox{$V_{\rm T} = -0.14$ m s$^{-1}$}; the
deceleration then amounts to 10.4~$\gamma$-units.  Thus, either of these
conditions suggests that $V_{\rm T}$  is more likely to be negative and,
accordingly, the deceleration probably exceeds 10~$\gamma$-units.

It turns out that there is an important relationship between the  deceleration
$\gamma$ of a companion and its lifetime against disintegration or longevity.
Based on their longevity, the companions are classified into three categories,
each described by a distinct $\gamma$ range (\mbox{Sekanina} 1982):\ persistent
companions (\mbox{$\gamma \!\leq\! 7 \: \gamma$-units}),~\mbox{short-lived}
\mbox{companions\,(\mbox{$20 \,\gamma$-units\,$\leq\!\gamma\!\leq\! 60\,\gamma$-units}), and$\:$minor$\:$com-} panions (\mbox{$\gamma \!>\! 60 \,\gamma$-units}).  

A measure of a companion's longevity is an {\it endurance\/}, $E$, defined as
a minimum lifetime against disintegration weighted by outgassing (Sekanina 1982)
\begin{equation}
E = \!\! \int_{t_{\rm frg}}^{t_{\rm fin}} \! \frac{dt}{r^2} = 1.015 \,
 [q(1 \!+\! e)]^{-\frac{1}{2}} (u_{\rm fin} \!-\! u_{\rm frg}),
\end{equation}
where $t_{\rm fin}$ is the time of last detection of the companion, $u_{\rm frg}$
and $u_{\rm fin}$ (in deg) the true anomalies at, respectively, $t_{\rm frg}$
and $t_{\rm fin}$, $r$ the heliocentric distance at time $t$, $q$ the
perihelion distance (in AU), and $e$ the eccentricity of the orbit.  The
endurance $E$ is then expressed in equivalent days at 1~AU from the Sun, or
{\it e-days\/}.  The demise of a companion is usually rather abrupt,
characterized by a sudden loss of the condensation, dramatic fading, and
rapid expansion and elongation of its apparent dimensions, all unquestionable
signs of the object's disintegration into a cloud of small fragments and dust.
As a rule, $t_{\rm fin}$ and $u_{\rm fin}$ are fairly well defined and so is $E$.

The available data suggest that the endurance correlates with the deceleration,
following closely an empirical law (Sekanina 1982)
\begin{equation}
E = C\,\gamma^{-0.4},
\end{equation}
where $C$ is a constant.  There appear to be three classes of companions, which
differ from one another by a~\mbox{degree} of brittleness (or cohesion); most
fragments (persistent, short-lived, and minor) of the split comets investigated
in the 1980s were members of the class of average brittleness, for which \mbox{$C =
200$ e-days}.  Among persistent companions there was, in addition, a class of
low brittleness (sturdier material), for which $C$ is about 4\,times higher,
while among short-lived and minor companions there also was a class of high
brittleness (more fragile material) with $C$ anout 2.3\,times lower.

In broad terms, the results in Table 6 and analysis of conditions (19) and (20)
suggest that the companion of C/2015~ER$_{61}$ was subjected to a moderate
deceleration that was close to an upper limit of the category of persistent
companions but not high enough to qualify as a short-lived companion.  However,
because of a small database available in the 1980s, the limits of the categories
were set rather arbitrarily.  Since then there have been several instances of
split comets with companions in the interim range between the persistent and
short-lived companions (\mbox{$7 \: \gamma$-units$\:< \! \gamma \! < 20 \:
\gamma$-units}).\footnote{A nice example is companion A of comet C/2001~A2, whose
separation from the primary nucleus B correlated temporally with Outburst~I (listed
here in Table~4).  The motion of this companion was affected by a deceleration
\mbox{$\gamma = 16.3 \pm 0.6 \:\gamma$-units} (Seka\-nina et al.\ 2002) and its
endurance was 63~e-days, in nearly perfect harmony with the law (22) for the
class of companions of average brittleness.  Other examples are the companion
to comet C/2005~A1 (LINEAR) with \mbox{$\gamma = 16.2 \pm 0.6 \: \gamma$-units}
(Seka\-nina 2005b); companions C (Solution III) and D$_2$ (Solution II) of comet
141P/Machholz with, respectively, \mbox{$\gamma = 11.5 \pm 2.5$} and \mbox{$15.9
\pm 4.8 \: \gamma$-units} (Sekanina 1999); and companions E, D, G, and B of comet
168P/Hergenrother, whose decelerations were in a range from \mbox{$9.2 \pm 0.3$}
to \mbox{$17.1 \pm 2.3 \; \gamma$-units}, although all four belonged to the
class of companions of high brittleness (the endurance ranging from 22 to
37~e-days; Sekanina 2010). See also Boehnhardt (2004).}

It is fair to say that the companion of C/2015~ER$_{61}$ was confined to a broad
boundary region between the persistent and short-lived companions.  The endurance
$E$ should unequivocally determine to  which brittleness class the companion
belongs.  The value of $E$ should also contribute to settling the issue of
correlation among the parameters of the fragmentation model.  Adopting for
this companion \mbox{$t_{\rm frg} \!-\! t_\pi = -34.7$ days} from Table~6
and \mbox{$t_{\rm fin} \!-\! t_\pi = +51.5$ days} (that is, \mbox{$t_{\rm fin}
= 2017$ June 30.4 UT}; Table~7), I find that \mbox{$u_{\rm fin} \!-\! u_{\rm
frg} = 98^{\circ}$} and \mbox{$E = 69$ e-days}.

\begin{figure*}
\vspace{-4.87cm}
\hspace{-0.86cm}
\centerline{
\scalebox{0.73}{
\includegraphics{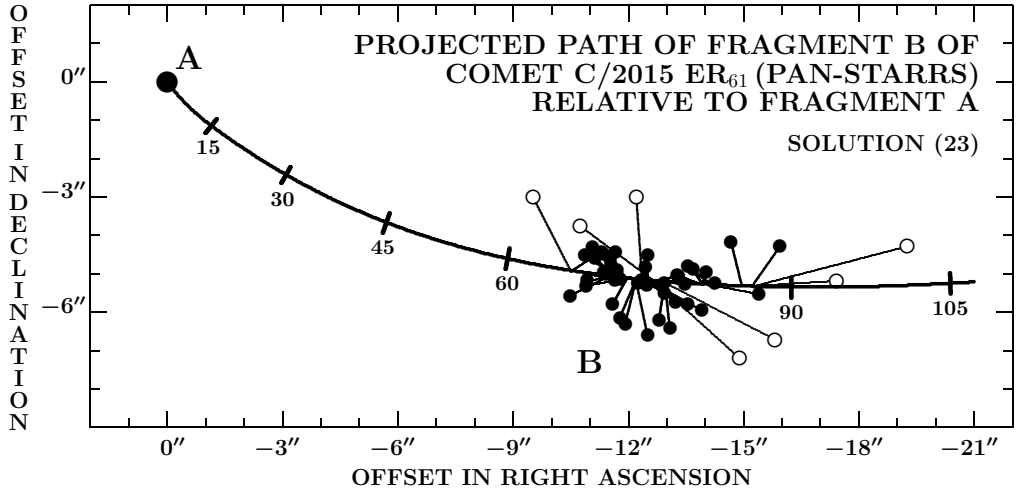}}}
\vspace{-10.3cm}
\caption{Path of the companion (fragment B) of C/2015 ER$_{61}$ relative to the
primary nucleus (fragment A) in projection onto the plane of the sky, as provided
by the solution (23) in the equatorial system of coordinates.  The solid circles
are the \mbox{B\,--\,A} offsets that were incorporated into the solution, the
open circles are those excluded because they left residuals exceeding
$\pm$1$^{\prime\prime\!\!}$.5 in one or both coordinates.  The numbers along
the path are days from fragmentation on 2017 April 5.2 UT (60~days refers to
June 4.2 UT, 90~days to July~4.2~UT,~etc.).{\vspace{0.5cm}}}  
\end{figure*}

One can readily rule out a chance of the companion belonging either to the class
of low brittleness~(requiring \mbox{$\gamma \sim 460 \: \gamma$-units}) or to the
class of high brittleness~(requiring \mbox{$\gamma \sim 1.8 \: \gamma$-units}).
For a member of the class of average brittleness, $E$ implies \mbox{$\gamma =
14.3 \: \gamma$-units}, well within the interval dictated by the condition
\mbox{$V_{\rm sep} \!<\! 1$~m~s$^{-1}$} coupled with Equation~(20).  From Equations
(19) it is now possible to derive the components of the separation velocity
and thus obtain a solution that (i)~fits the \mbox{B\,--\,A} offsets;
(ii)~complies with the condition that the fragmentation event closely correlated
with the outburst; (iii)~satisfies the condition of low separation velocity;
and (iv)~is consistent with the correlation between the deceleration and
endurance of the companion.  The parameters and other data describing this
final solution are:
\begin{eqnarray}
t_{\rm frg} & = & 2017 \:{\rm April \,\, 5.2 \,\, UT}, \nonumber \\
t_{\rm frg} \!-\! t_\pi & = & -34.7\:{\rm days}, \nonumber \\
\gamma & = & 14.3 \:\gamma\mbox{-units}, \nonumber \\[-0.05cm]
V_{\rm sep} & = & 0.90 \:{\rm m\,s}^{-1} \!, \nonumber \\[-0.05cm]
V_{\rm R} & = & +0.38 \:{\rm m\,s}^{-1} \!, \nonumber \\[-0.05cm]
V_{\rm T} & = & -0.69 \:{\rm m\,s}^{-1} \!, \nonumber \\[-0.05cm]
V_{\rm N} & = & -0.44 \:{\rm m\,s}^{-1} \!, \nonumber \\
E & = & 69 \:\mbox{e-days}, \nonumber \\
\mbox{Mean\,\,{\small \it O--C}}\, & = & \pm 0^{\prime\prime\!\!}.65,
 \nonumber \\[-0.05cm]
\Sigma\,\mbox{\small ($\!${\it O--C\/})}^2 & = & 34.675 \:{\rm arcsec}^2, \nonumber \\
\mbox{Total Dataset} & = & \mbox{49 offsets}, \nonumber \\
\mbox{Dataset Used} & = & \mbox{42 offsets}, \nonumber \\
\mbox{Time Span Covered} & = & \mbox{2017$\:$June$\:$11.4--30.4$\:$UT}.
\end{eqnarray}

To complete the portrayal of the companion's birth and orbital evolution in
terms of the solution (23), I plot in Figure~8 its resulting path relative to the
primary nucleus and list in Table~7 the 49~offsets and their ``observed minus
computed,'' or {\it O--C}, residuals from the solution.  Each offset entry in
this table is a difference between a single astrometric position of the companion
and an average of up to 12 motion-corrected astrometric positions of the primary
nucleus at times of up to 0.01~day apart.  An internal error of these averaged
offsets is typically $\pm$0$^{\prime\prime\!\!}$.4 in right ascension and
$\pm$0$^{\prime\prime\!\!}$.3 in declination.  Seven of the 49 tabulated offsets
were excluded from the solution because they left residuals (parenthesized in
Table~7) in excess of $\pm$1$^{\prime\prime\!\!}$.5 in one or both coordinates.
The table shows that the distribution of residuals of the included offsets is
satisfactory in that it exhibits no systematic trends in either right ascension
or declination.

\begin{table}[b]
\vspace{-3.8cm} 
\hspace{4.22cm}
\centerline{
\scalebox{1}{
\includegraphics{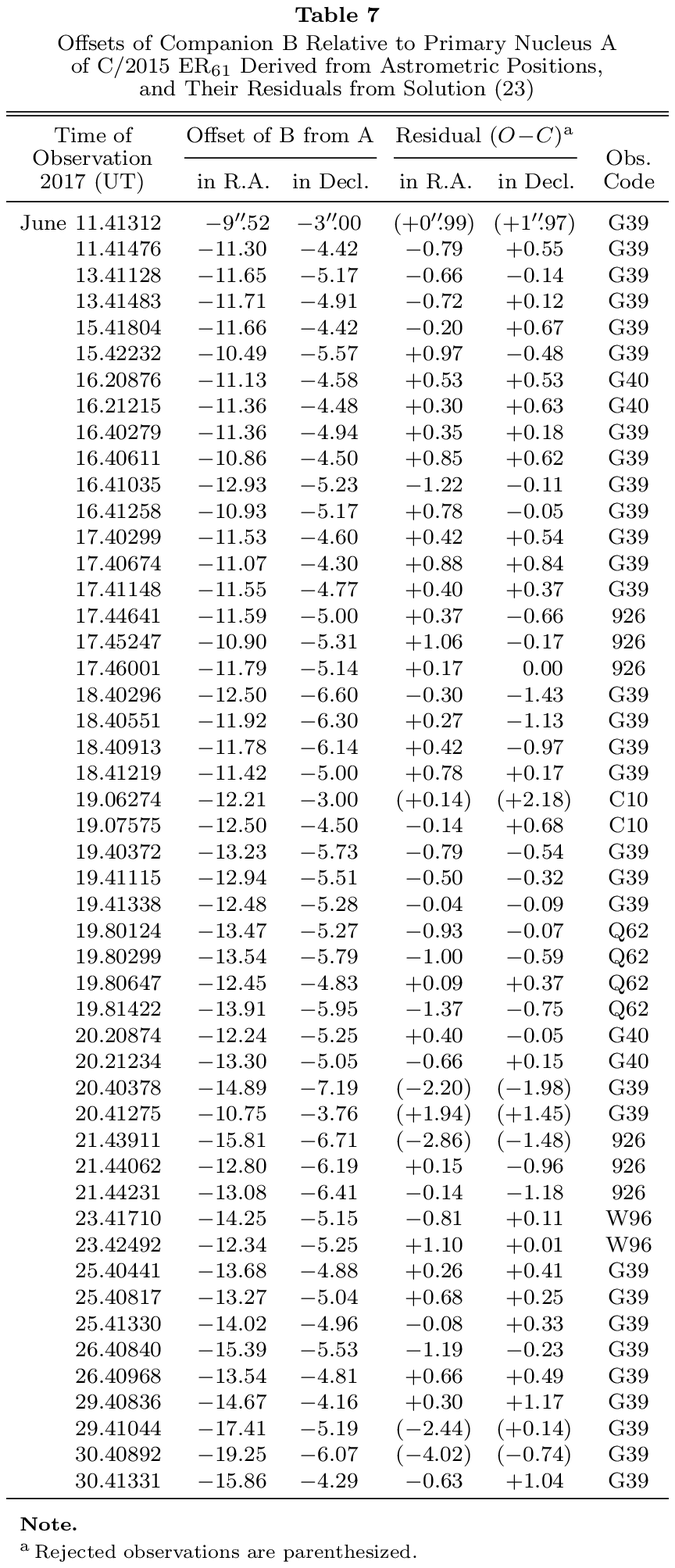}}}
\vspace{-5.4cm}
\end{table}

As a final comment on the companion's relative motion, I should mention that
Bachini (2017), of Stazione Astronomica BS-CR, Santa Maria a Monte, Italy,
observing with a 25-cm f/10 Schmidt-Cassegrain with an f/6.3 focal reducer
(Code K47), is reported as having provided a set of four astrometric observations
of the companion, spanning 17~minutes on 2017 August 23.\footnote{Because of the
brevity of recorded astrometric observations in the {\it Minor Planet Circulars\/},
one cannot rule out a possibility that the positions were attributed to the
companion (rather than to the primary) by the {\it Minor Planet Center\/}'s
staff, not by the observer.}~Four~circum\-stances that make this
identification highly suspect include:\ (i)~a nuclear magnitude of 14.6, which
is comparable to the brightness of the primary nucleus reported at the time by
other observers; (ii)~a gap of nearly 8~weeks from the previous observation of
the companion; (iii)~no simultaneous astrometry of the primary; and (iv)~the
orbital elements for the companion from a solution that incorporated these
dubious astrometric positions ({\it Minor Planet Center\/} 2017) being listed
with no mean residual of the fit, which is extremely unusual.  Because of (iii),
one cannot compute the \mbox{B\,--\,A} offset, but the residuals from the orbit
of the primary are a fair proxy~for~the missing offset.  They average
+21$^{\prime\prime\!}$ in right ascension and +3$^{\prime\prime\!}$ in
declination.  The solution (23) predicts for August~23 the companion's offset
of $-$32$^{\prime\prime\!}$ in right ascension and $-$5$^{\prime\prime\!}$ in
declination, leaving enormous residuals of +53$^{\prime\prime\!}$ and
+8$^{\prime \prime\!}$ in the two coordinates, respectively.  When the
August~23 proxy offset was incorporated into the input dataset, the resulting
solution failed miserably, leaving strong systematic residuals of several
arcsec.  There is no doubt whatsoever that the August~23 astrometric positions
reported from Code K47 refer to the primary nucleus, but are off,
especially in right ascension.  As a result, the published parabolic orbital
elements for the companion from a solution that incorporated these positions
({\it Minor Planet Center\/} 2017) are meaningless.

\subsection{Companion's Light Curve}
If the comet's nucleus split in the course of the outburst, then the question
of why it took more than nine weeks to detect the companion for the first time
needs to be answered.  While cometary companions are never recognized by a
terrestrial observer soon after the breakup because of their proximity to
the primary, the proposed solution (23) suggests that the separation between
the fragments was 2$^{\prime\prime}$ on April~22, 17.6~days after breakup;
4$^{\prime\prime}$ on May~5, 30.3~days after breakup; and 6$^{\prime\prime}$
on May~16, 41~days after breakup.  Thus, insufficient separation distance is
unlikely to provide an explanation for the considerable delay, and the apparent
culprit may instead be the companion's excessive faintness.  In a search for
evidence of the suspected cause, I compiled and normalized the companion's
CCD nuclear magnitudes and compared the resulting light curve with that of
the primary nucleus obtained from the same data sources.  The results turned
out to be quite compelling.

\begin{figure}[b]
\vspace{-2.7cm}
\hspace{0.85cm}
\centerline{
\scalebox{0.725}{
\includegraphics{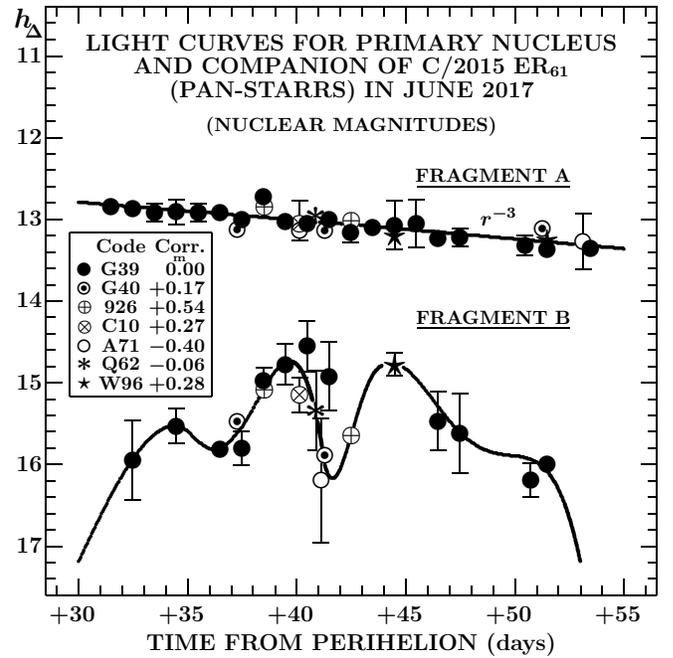}}}
\vspace{-9.9cm}
\caption{Light curves of the companion (fragment B; bottom) and the primary
nucleus (fragment A; top) of comet C/2015~ER$_{61}$ over the entire period
of time in June 2017 when the companion was under observation.  The plotted
CCD nuclear magnitudes $h_\Delta$ have been normalized to a unit geocentric
distance by an inverse square power law.  Although no phase correction was
applied, the model by Marcus (2007), based on a compound Henyey-Greenstein
dust scattering law, shows that introducing the correction would result in
relative changes not exceeding $\pm$0.02~mag (merely shifting the plot
up by almost 1~mag) because of a limited range of phase angles.  The plot is
based on 167 individual nuclear-magnitude determinations for the primary nucleus
and 52 such determinations for the companion.  They were measured from images
taken in rapid succession over less than $\sim$20~minutes and averaged into
31 and 20~plotted data points for, respectively, the primary nucleus and the
companion, with the mean short-term scatter depicted by the error bars.  The
entries without an error bar have a mean error not exceeding $\pm$0.1~mag,
the size of the used symbols.  The magnitudes were provided along with
astrometric positions by the six sources listed in Table~5 plus by code A71
(Stixendorf, Austria, 30-cm f/4 reflector, observers M.~J\"{a}ger et al.);
the data are available from the {\it Minor Planet Center\/}'s database.  The
fairly smooth light curve of the primary nucleus was used to compute
systematic corrections for the observing codes to convert the magnitudes
of both fragments to the brightness scale of code G39; the corrections are
in~the~inset.{\vspace{0cm}}} 
\end{figure}

The data points, plotted in Figure 9, are averages from the nuclear magnitudes
reported by the six sources listed in Table~5 and by the observers at Stixendorf,
Austria (code A71; see the caption).  The individual magnitudes were measured
by the observers in images taken in rapid succession covering periods of up
to $\sim$20~minutes (for the reference, see footnote 2).  Each average has an
internal mean error that measures a combined effect of observational uncertainties
and genuine brightness fluctuations on a scale of a fraction of one hour.  The
figure depicts dramatic differences between the photometric behavior of the
primary and the companion on several time scales:

(1) On a scale comparable to the span of time covered in Figure~9, of approximately
three weeks; the primary nucleus was fading fairly smoothly at a rate that closely
followed an inverse cube power law of heliocentric distance, whereas the companion
was brightening at the beginning but fading at the end at a rapid rate, suggesting
that its excessive faintness prior to June~11 was indeed the reason for its
nondetection.

(2) On a scale of several days, the brightness variations of the primary nucleus 
were confined to a rate~of~seldom higher than 0.1~mag, whereas the amplitude of
the companion's brightness fluctuations clearly exceeded 1~mag.  The curve fitted
to these data points is only a crude approximation to the irregular brightness
variations.

(3) The primary nucleus is seen in Figure 9 to have been imaged and measured on
many more days than the companion; a possible reason for this disparity could
be an intermittent excessive faintness of the companion on the days from which
its astrometry is unavailable, providing further potential evidence for its
brightness fluctuations on a scale of days.

(4) Although one can expect instrumental errors in determining a magnitude to
increase somewhat with decreasing brightness, the difference between the
large error bars of the magnitude measurements of the companion compared to
the primary nucleus is too striking to be explained by observational
uncertainties.  The highly unequal error bars imply that the primary nucleus
and the companion differed significantly in terms of their brightness
fluctuations on a scale of a fraction of one hour.

In summary, I find that a major disparity in the behavior of the two
nuclear fragments and a considerably larger amplitude of the companion's
brightness fluctuations took place over intervals of time that spanned at
least three orders of magnitude.  The extreme variability of the companion's
activity appears to have been associated with the object's progressive
fragmentation that probably resulted in its complete disintegration by the
end of June or in the first days of July. Paradoxically, it was this crumbling
process that made it possible to detect the companion in the first place;
before these developments came about in earnest, we had been unaware of
the companion's existence.

Highly irregular variations in the brightness of companion nuclei were
repeatedly documented in numerous past investigations of the split comets;
the light curves exhibited by fragments of the long-period comets
C/1947~X1, C/1956~F1, and C/1975~V1, see Sekanina (1982); of 73P, an
example of the short-period comets, see Boehnhardt (2004), Sekanina (2008a).

\section{Conclusions}
Two striking phenomena displayed by C/2015~ER$_{61}$, a dust-poor comet, were
an outburst that peaked nearly five weeks before perihelion and a double nucleus
that was first noticed about one month after perihelion and was followed for
nearly three weeks.  Three major objectives addressed in this paper are:\
(1)~the onset and physical evolution of the outburst; (2)~the conditions at
birth of the companion nucleus B and its orbital and physical evolution in
comparison with the primary nucleus A; and, most importantly, (3)~a potential
close relationship between the outburst and nuclear splitting, in spite of a
gap of more than nine weeks between them.{\hspace{0.2cm}}

The outburst is detected in, and its timing constrained by, five
datasets:\ (i)~a visual light curve; (ii)~a light curve based on selected
CCD total magnitudes; (iii)~a light curve based on selected CCD nuclear
magnitudes; (iv)~a large-scale imaging; and (v)~a curve of the {\it Af}$\rho$
parameter, which is a proxy of the comet's dust-production rate.  Four of
these sets --- (i) and (iii) through (v) --- turn out to be broadly consistent
with the model adopted in this paper:\ an onset of the outburst 36.0~days
before perihelion, on 2017 April~3.9~UT, and a sharp, spike-like peak (typical
for dust-poor comets) some 2.6~days later.  The visual light-curve data
are distributed densely enough that the uncertainty of the onset time can be
estimated at as little as $\pm$0.1~day and the time of peak brightness at
$\pm$1~day.  The light curve based on the selected CCD total magnitudes
implies instead March~28.6~UT for the onset time, with an estimated
uncertainty of $\pm$1~day.  The two critical CCD observations that appear to
rule out a later onset time are by \mbox{T.~Takahashi} on March~30.8~UT and
by H.~Abe on April~1.8~UT (Figure~1), both showing the comet's brightness
to have become increasingly elevated compared to the level before March~20.
The other datasets do not suggest any such effect; the reason for this
discrepancy remains unclear.

The rise time, defined as a temporal distance from~the onset to the peak
brightness, measures the active phase of the outburst.  Comparison of
C/2015~ER$_{61}$~with~some other flaring-up comets suggests that the
active~phase~is usually brief, not exceeding a few days, although~there are
exceptions, as illustrated by comet~73P~in~1995~or~by Event~III of
C/2001~A2 (Table~4).  The active~phase~of 2.6$\,\pm\,$1~days for the
adopted outburst model of C/2015 ER$_{61}$ thus appears to be of a
rather typical duration.

Investigated in relation to the primary nucleus A, the motion of the companion B
is based on 42 out of a total of 49 \mbox{B\,--\,A} offsets in right ascension
and declination, which are fitted by a multiparameter model for the split comets
(Sekanina 1978, 1982), extensively applied in the past.  Some of the solutions
obtained for C/2015~ER$_{61}$, presented in Table~6, suggest that the
companion's motion was subjected to a moderate nongravitational deceleration
and that the time of the companion's separation from the parent nucleus is
rather poorly determined (and highly correlated with the radial component
of the separation velocity), but is most probably between 50$\,\pm\,$8 and
27$\,\pm\,$9~days before perihelion, or, nominally, between April~21 and
May~13, thus bracketing the outburst whose active phase is centered on
34.7$\,\pm\,$0.5~days before perihelion, or, nominally, on April~5.2~UT.  With
the high probability that the outburst was a product and signature of nuclear
fragmentation, a three-parameter model from Table~6 was further refined by
incorporating a relationship between the companion's longevity, or endurance,
and nongravitational deceleration for cometary fragments of medium brittleness
(Sekanina 1982).\footnote{More recently, Boehnhardt (2004) suggested an alternative,
averaged fit to an expanded dataset; disregarding the previously introduced
classes of fragment brittleness, he finds a much steeper dependence of the
endurance on the nongravitational deceleration, even though his plot still
shows that the variations among both the most and the least enduring fragments
follow a law that is substantially flatter.  Application of Boehnhardt's
relationship between the endurance and deceleration to C/2015~ER$_{61}$ would
result in a solution for the companion that is outside a range judged as
plausible by other applied constraints (Section~5.6).{\vspace{-0.2cm}}}

The adopted fragmentation model --- solution (23) --- indicates that the
companion separated from the parent nucleus with a velocity of 0.9~m~s$^{-1}$
in a direction generally away from the Sun, opposite to the comet's orbital
motion, and below the orbital plane.  The deceleration, 14.3 units of 10$^{-5}$
the Sun's gravitational acceleration, offers a crude estimate of less than one
hundred meters for the overall dimensions of the companion, too small to
detect unless in a state of advanced fragmentation.

This implication appears to be corroborated by observational evidence.
Contrary to the primary nucleus, the companion displayed major fluctuations
in brightness on time scales that spanned at least three orders of magnitude,
from a fraction of one hour to weeks.  The more-than-nine-week gap between the
companion's birth and its first detection appears to be a corollary of this
brightness variability.  Specifically, the companion may have been~too~faint
for detection before June~11, a distinct possibility supported by a rapid
brightening at the time (Figure~9).  Indeed, it apparently was the debris~the
companion was shedding and activity derived from newly exposed areas of
volatile substances --- products of progressive fragmentation --- that greatly
facilitated the detection.  Ironically, this same process resulted in the
companion's impending disintegration by the end of June or in early July,
also in line with the light curve.

The story of an outburst and ensuing nuclear~\mbox{duplicity} of comet
C/2015 ER$_{61}$ appears to be rather reminiscent of similar phenomena
observed in other flaring-up and split comets.~In-depth comparison of this
object's~outburst with such events in a number of past comets was employed
to introduce a new set of outburst parameters, while the data on the
companion were exploited to offer a comprehensive model for its orbital motion
and to investigate an interaction between the companion's orbital and physical
properties.  The author hopes that, jointly with his recent work on comet
168P/Hergenrother, this investigation contributes both to gaining greater~insight
into the behavior of the split comets and to the understanding of the physical
relationship between outbursts and nuclear fragmentation.\\[-0.1cm]

This research was carried out at the Jet Propulsion Laboratory, California
Institute of Technology, under contract with the National Aeronautics and
Space~Administration.\\[-0.2cm]
\begin{center}
{\footnotesize REFERENCES}
\end{center}
\vspace*{-0.3cm}
\begin{description}
{\footnotesize
\item[\hspace{-0.3cm}]
A'Hearn, M. F., Millis, R. L., Schleicher, D. G, et al. 1995, Icarus,{\linebreak}
 {\hspace*{-0.6cm}}118, 223
\\[-0.57cm]
\item[\hspace{-0.3cm}]
Bachini, M. 2017, MPC 105599
\\[-0.57cm]
%
%
\item[\hspace{-0.3cm}]
Bessell, M. S. 1979, PASP, 91, 589
\\[-0.57cm]
\item[\hspace{-0.3cm}]
Beyer, M. 1962, Astron. Nachr., 286, 219
\\[-0.57cm]
\item[\hspace{-0.3cm}]
Bobrovnikoff, N. T. 1927, ApJ, 66, 439
\\[-0.57cm]
\item[\hspace{-0.3cm}]
Bobrovnikoff, N. T. 1928, PASP, 40, 164
\\[-0.57cm]
\item[\hspace{-0.3cm}]
Boehnhardt, H.\ 2004, in Comets II, ed.\ M.\ C.\ Festou, H.\ U.\ Keller,{\linebreak}
 {\hspace*{-0.6cm}}\& H.\ A.\ Weaver (Tucson, AZ:\ University of Arizona), 301
\\[-0.57cm]
\item[\hspace{-0.3cm}]
Gibson, B., Goggia, T., Primak, N., et al.\ 2015, MPEC 2015-F124
\\[-0.57cm]
\item[\hspace{-0.3cm}]
Gibson, B., Goggia, T., Primak, N., et al.\ 2016, MPC 97628
\\[-0.57cm]
\item[\hspace{-0.3cm}]
Green, D.\,W.\,E. 2016, CBET 4249
\\[-0.57cm]
\item[\hspace{-0.3cm}]
Green, D.\,W.\,E. 2017a, CBET 4383
\\[-0.57cm]
\item[\hspace{-0.3cm}]
Green, D.\,W.\,E. 2017b, CBET 4409
\\[-0.57cm]
\item[\hspace{-0.3cm}]
Hambsch, F.-J., \& Bryssinck, E. 2017, MPEC 2017-N28
\\[-0.57cm]
\item[\hspace{-0.3cm}]
Hambsch, F.-J., Bryssinck, E., \& Hambsch, J. 2017a, MPEC 2017-{\linebreak}
 {\hspace*{-0.6cm}}M09
\\[-0.57cm]
\item[\hspace{-0.3cm}]
Hambsch, F.-J., Bryssinck, E., \& Hambsch, J. 2017b, MPEC 2017-{\linebreak}
 {\hspace*{-0.6cm}}N58
\\[-0.35cm]
\item[\hspace{-0.3cm}]
Hosek, M. W. Jr., Blaauw, R. C., Cooke, W. J., et al.\ 2013,~AJ,~145,{\linebreak}
 {\hspace*{-0.6cm}}122
\\[-0.57cm]
\item[\hspace{-0.3cm}]
King, B. 2017, {\tt http://www.skyandtelescope.com/astronomy-news{\linebreak}
 {\hspace*{-0.6cm}}/comet-er61-panstarrs-in-outburst-binocular-bright}
\\[-0.57cm]
\item[\hspace{-0.3cm}]
Marcus, J. N. 2007, ICQ, 29, 39
\\[-0.57cm]
%
%
%
%
\item[\hspace{-0.3cm}]
Masi, G., \& Schwartz, M. 2017, MPEC 2017-M09
\\[-0.57cm]
\item[\hspace{-0.3cm}]
Miles, R., Faillace, G. A., Mottola, S., et al.\ 2016, Icarus, 272, 327
\\[-0.57cm]
\item[\hspace{-0.3cm}]
{\it Minor Planet Center\/} 2015, MPEC 2015-F124
\\[-0.57cm]
\item[\hspace{-0.3cm}]
{\it Minor Planet Center\/} 2017, MPC 106344
\\[-0.57cm]
\item[\hspace{-0.3cm}]
M\"{u}ller, G. 1884a, Astron. Nachr., 107, 381
\\[-0.57cm]
\item[\hspace{-0.3cm}]
M\"{u}ller, G. 1884b, Astron. Nachr., 108, 161
\\[-0.57cm]
\item[\hspace{-0.3cm}]
Nakano, S. 2017, Nakano Note 3379
\\[-0.57cm]
\item[\hspace{-0.3cm}]
Primak, N., Schultz, A., Watters, S., et al.\ 2016, MPEC 2016-C01
\\[-0.57cm]
\item[\hspace{-0.3cm}]
Roemer, E. 1959, PASP, 71, 546
\\[-0.57cm]
\item[\hspace{-0.3cm}]
Sekanina, Z. 1978, Icarus, 33, 173
\\[-0.57cm]
\item[\hspace{-0.3cm}]
Sekanina, Z. 1982, in Comets, ed. L. L. Wilkening (Tucson, AZ:
{\hspace*{-0.6cm}}University of Arizona), 251
\\[-0.57cm]
\item[\hspace{-0.3cm}]
Sekanina, Z. 1999, A\&A, 342, 285
\\[-0.57cm]
\item[\hspace{-0.3cm}]
Sekanina, Z. 2000, IAUC 7471
\\[-0.57cm]
\item[\hspace{-0.3cm}]
Sekanina, Z. 2005a, Int. Comet Quart., 27, 225
\\[-0.57cm]
\item[\hspace{-0.3cm}]
Sekanina, Z. 2005b, IAUC 8562
\\[-0.57cm]
\item[\hspace{-0.3cm}]
Sekanina, Z. 2008a, Int. Comet Quart., 30, 3
\\[-0.57cm]
\item[\hspace{-0.3cm}]
Sekanina, Z. 2008b, Int. Comet Quart., 30, 63
\\[-0.57cm]
\item[\hspace{-0.3cm}]
Sekanina, Z. 2009, Int. Comet Quart., 31, 5
\\[-0.57cm]
\item[\hspace{-0.3cm}]
Sekanina, Z. 2010, Int. Comet Quart., 32, 45
\\[-0.57cm]
\item[\hspace{-0.3cm}]
Sekanina, Z. 2012, {\it Giant\,Explosions,\,Cascading\,Fragmentation,\,and{\linebreak}
{\hspace*{-0.6cm}}Episodic Aging of Comets\/}. Academy of Sciences of the
Czech~Re-{\linebreak}
{\hspace*{-0.6cm}}public, Praha, 184\,pp.\ (In Czech.)
\\[-0.57cm]
\item[\hspace{-0.3cm}]
Sekanina, Z., \& Kracht, R. 2016, ApJ, 823, 2
\\[-0.57cm]
\item[\hspace{-0.3cm}]
Sekanina, Z., Jehin, E., Boehnhardt, H., et al.\ 2002, ApJ, 572, 679
\\[-0.57cm]
\item[\hspace{-0.3cm}]
Trigo-Rodr\'{\i}guez,\,J.\,M., Garc\'{\i}a-Hern\'andez,\,D.\,A., S\'anchez,\,A.,~et~al.{\linebreak}
 {\hspace*{-0.6cm}}2010, MNRAS, 409, 1682
\\[-0.66cm]
\item[\hspace{-0.3cm}]
Vogel, H. C. 1884, Astron. Nachr., 131, 373}
\vspace{-0.23cm}
\end{description}
\end{document}